\documentclass[12pt]{iopart}
\usepackage{graphicx}
\usepackage{bm}
\usepackage{amsgen,amssymb,amsfonts,amsbsy}
\newcommand{\NullNormal}{k}         
\newcommand{\NullExpansion}{\theta} 
\newcommand{\NullShear}{\sigma}     
\newcommand{\FourCD}{{^{(4)}\nabla}}  

\newcommand{\Eadm}{M_{\rm ADM}}
\newcommand{\td}{{\rm d}}    
\newcommand{\SMetric}{\gamma}     
\newcommand{\Lapse}{\alpha}       
\newcommand{\Shift}{\beta}   

\newcommand{\ExCurv}{K}      
\newcommand{\TrExCurv}{K}    
\newcommand{\A}{A}           

\newcommand{\dtime}{\partial_t}   
\newcommand{\SCD}{{\nabla\!}}     
\newcommand{\SLong}[1]{({\mathbb L}{#1})} 

\newcommand{\SFlatMetric}{f}      

\newcommand{\STwoMetric}{h}         
\newcommand{\SSpatialNormal}{s} 

\newcommand{\CF}{\psi}              

\newcommand{\CMetric}{{\tilde{\gamma}}}    
\newcommand{\CLapse}{\tilde{\alpha}}     
\newcommand{\dtCMetric}{\tilde{u}}  

\newcommand{\CA}{\tilde{A}}         
\newcommand{\CRicci}{\tilde{R}}     
\newcommand{\CRicciS}{\tilde{R}}    

\newcommand{\CCD}{{\tilde\nabla}\!} 
\newcommand{\CCDu}{{\tilde\nabla}}  
\newcommand{\CLong}[1]{(\tilde{\mathbb L}{#1})} 

\newcommand{\CTwoMetric}{\tilde{h}}         
\newcommand{\CSpatialNormal}{\tilde{s}}

\newcommand{\OmegaOrbitID}{\Omega_0}  

\newcommand{\FittingFactor}{\mu}
\newcommand{\xirot}{\xi_{\rm rot}}
\newcommand{\xiinsp}{\xi_{\rm insp}}

\newcommand{\PropSep}{s}  
\newcommand{\OmegaOrbitEv}{\omega}  

\begin{document}
\vspace{-2.5cm} 

\title{Reducing orbital eccentricity in binary black hole simulations}

\author{Harald P. Pfeiffer$^1$, Duncan A. Brown$^{1,2}$, Lawrence E. Kidder$^3$, Lee Lindblom$^1$, Geoffrey Lovelace$^1$, and Mark A. Scheel$^1$}

\address{${}^1$Theoretical Astrophysics, California Institute of
Technology, Pasadena, California 91125}

\address{${}^2$LIGO Laboratory, California Institute of Technology,
  Pasadena, California 91125}

\address{${}^3$Center for Radiophysics and Space Research, 
Cornell University, Ithaca, New York, 14853}

\date{\today}

\begin{abstract}
Binary black hole simulations starting from quasi-circular ({\it
i.e.,\/} zero radial velocity) initial data have orbits with small
but non-zero orbital eccentricities.
In this paper the quasi-equilibrium initial-data method is
extended to allow non-zero radial velocities to be specified 
in binary
black hole initial data.  New low-eccentricity
initial data are obtained by adjusting the orbital frequency and
radial velocities to minimize the orbital eccentricity, and 
the resulting ($\sim 5$ orbit) evolutions are compared
with those of quasi-circular initial data.
Evolutions of the quasi-circular data clearly
show eccentric orbits, with eccentricity that decays over
time.  The precise decay rate depends on the definition of
eccentricity; if defined in terms of variations in the orbital frequency, 
the decay
rate agrees well with the prediction of Peters (1964).
The gravitational waveforms, which contain $\sim 8$ cycles in the
dominant $l=m=2$ mode, are largely unaffected by the eccentricity
of the quasi-circular initial data. The overlap between the dominant
mode in the quasi-circular evolution and the same mode in the
low-eccentricity evolution is about $0.99$.
\end{abstract}

\pacs{04.25.Dm, 04.30.Db, 04.70.Bw}

\submitto{\CQG}

\section{Introduction}
\label{s:Introduction}

The inspiral and merger of binary black holes is one of the most promising
sources for current and future generations of interferometric gravitational
wave detectors such as LIGO and VIRGO~\cite{Barish:1999,Acernese:2002}. 
The initial LIGO detectors,
which are currently operating at design sensitivity, could detect binary black
hole inspirals up to distances of several hundred megaparsecs. In order to take
full advantage of the sensitivity of these detectors, detailed knowledge of the
gravitational waveform is required.

Recent breakthroughs in numerical relativity have allowed several research
groups to simulate binary black hole inspirals for multiple
orbits~\cite{Pretorius2005a,Baker2006a, Campanelli2006a,%
Scheel2006,Bruegmann2006}.  Because of the large computational cost of these
simulations, only a small number of orbits can be followed. Therefore
it is important to begin these simulations with initial data that closely
approximate
a snapshot of a binary black hole system that is only a few orbits from merger.
During the inspiral, the orbits of binary compact objects circularize
via the emission of gravitational
waves~\cite{Peters1964}, so binaries formed from stellar evolution
(rather than dynamical capture) are expected to have very small
eccentricities by the time they enter the sensitive band of ground
based detectors.  Because of this, the assumption of a
quasi-circular orbit ({\it i.e.,\/} zero radial velocity) 
has been widely used in the construction of
binary black hole initial
data~\cite{Cook1994,Baumgarte2000,Gourgoulhon2001,Grandclement2001,%
Baker2002b,Cook2002,Tichy2002,Pfeiffer2003a,Cook2004,Yo2004,Tichy2004,%
Caudill-etal:2006,Dennison2006,Yunes2006a,Yunes2006b}.  Specifically,
quasi-equilibrium data~\cite{Cook2004} and the
``QC-sequence''~\cite{Baker2002} of puncture data~\cite{Brandt1997}
seem to be the most popular, and both of these assume a quasi-circular orbit.
However, inspiraling compact objects have a small inward radial velocity,
and neglecting this velocity when constructing initial data will
lead to eccentricity in the subsequent evolution, as discussed
in the context of post-Newtonian theory in Ref.~\cite{Miller2004}, and
found numerically in Refs.~\cite{Buonanno-Cook-Pretorius:2006}.

\begin{figure}
\centerline{\includegraphics[width=5in]{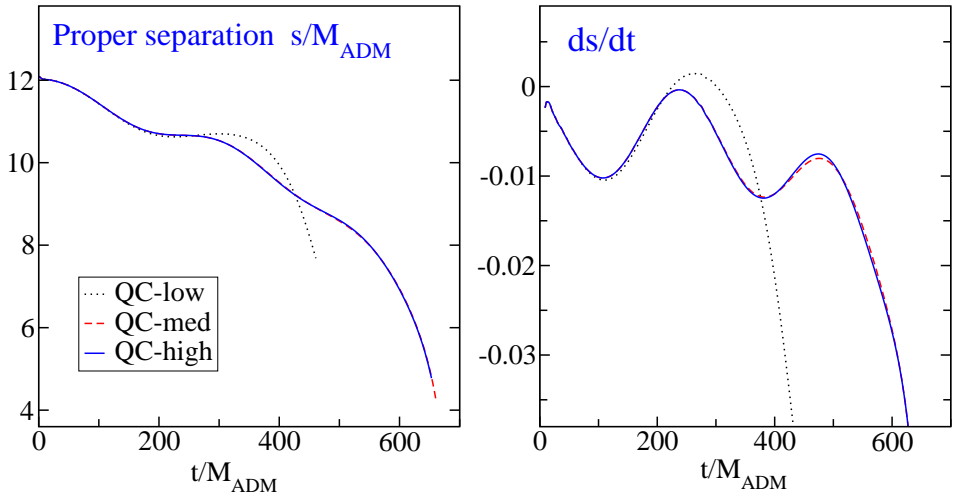}
}
\caption{\label{fig:EllipticExample}Evolution of quasi-circular
  initial data.  The left panel shows the proper separation $\PropSep$
  between the apparent horizons, computed at constant coordinate time
  along the coordinate line connecting the centers of the horizons,
  and the right panel shows
  its time derivative $d\PropSep/dt$.  This evolution was run at three
  different resolutions, with the medium and high resolution tracking
  each other very closely through the run.
  }
\end{figure}

The Caltech/Cornell collaboration has recently completed successful
long-term simulations of inspiraling binary black
holes~\cite{Scheel2006} using a pseudo-spectral multi-domain method.
This technique was used to evolve a particular quasi-circular quasi-equilibrium
binary black hole initial data set 
(coordinate separation $d=20$ from Table IV of
Ref.~\cite{Cook2004}).  \Fref{fig:EllipticExample} shows the proper
separation $\PropSep$ between the horizons and the radial velocity
$d\PropSep/dt$ as functions of time for this evolution.  The rapid
convergence afforded by spectral methods is apparent; the medium and
high resolutions are nearly indistinguishable on the plot.
Eccentricity of the orbit in the form of oscillatory variations in $\PropSep$ and $d\PropSep/dt$ is, unfortunately, also clearly apparent.

This noticeable eccentricity suggests two questions: First, how can
initial data with the appropriate black hole radial velocities 
be constructed for non-eccentric inspirals?  
Second, how do evolutions
of quasi-circular initial data differ from those of non-eccentric
initial data?  This paper addresses both questions.  In
Sec.~\ref{sec:2}, we incorporate nonzero radial velocities into the
quasi-equilibrium method to construct binary black hole initial data.  This results
in one additional parameter for equal mass initial data, the radial velocity
$v_r$.  Section~\ref{sec:numerical-methods} briefly discusses our
numerical methods.  Section~\ref{sec:3} describes how we choose $v_r$
and the the orbital frequency $\OmegaOrbitID$ for equal mass
co-rotating binary black holes, and presents numerical evolutions
of the resulting low-eccentricity initial data.  This section also presents
convergence tests of these binary black hole evolutions;
we examine both convergence with respect to spatial resolution and convergence
with respect to the radius of the outer boundary of the computational domain. 
Section~\ref{sec:4}
examines the differences between evolutions of quasi-circular initial
data and low-eccentricity initial data.  We close with a summary and 
discussion of these results in Sec.~\ref{sec:5}.

\section{Quasi-equilibrium data with nonzero radial velocity}
\label{sec:2}

In this section we extend the quasi-equilibrium
approach~\cite{Cook2002, Pfeiffer2003a,Cook2004, Caudill-etal:2006} to
allow specification of nonzero radial velocities of the black holes.  We
proceed in three steps: First, we summarize the 
construction of quasi-equilibrium data using co-rotating
coordinates~\cite{Cook2004, Caudill-etal:2006}.  Second, we show that
the {\em identical} quasi-circular initial data can be obtained by
solving essentially the same equations but in an asymptotically
inertial coordinate system; the major difference is that one must
require the black holes to {\em move} on circular trajectories, rather
than remaining fixed in the coordinate system.  Third, we generalize 
from black holes moving on circular trajectories to black holes moving
on inspiral trajectories.

\subsection{Overview}

We use the nomenclature of Ref.~\cite{Cook2004}; the spacetime line element
is written in the usual $3\!+\!1$-form,
\begin{equation}
\td s^2
=-\Lapse^2 \td t^2+\SMetric_{ij}(\td x^i+\Shift^i\td t)(\td x^j+\Shift^j\td t),
\end{equation}
where $\SMetric_{ij}$ is the 3-metric induced on a $t={\rm constant}$ spatial
hypersurface, $\Lapse$ is the lapse function and $\Shift^i$ is
the shift vector.  Latin indices label spatial coordinates, and Greek
indices label spacetime coordinates.  The extrinsic curvature of the
hypersurface is defined by
\begin{equation}\label{eq:DefExCurv}
\ExCurv_{\mu\nu}\equiv - \SMetric_{\mu}{}^\rho\SMetric_\nu{}^\sigma\; \FourCD_{(\rho}n_{\sigma)},
\end{equation}
where $\FourCD$ is the spacetime derivative operator and $n_\mu$ is
the future-pointing unit normal to the slice\footnote{Since
  $\ExCurv_{\mu\nu}$ is a spatial tensor, $\ExCurv_{\mu\nu}n^\nu=0$, 
  its spatial components
  $K_{ij}$ carry all its information.  Almost all tensors in this paper are
  spatial, and we use spatial indices here whenever possible.  }.  We
use the extended conformal thin sandwich
formalism~\cite{York1999,Pfeiffer2003b} to construct constraint-satisfying 
initial data.  In this approach, the three-dimensional metric
is spit into a conformal metric $\CMetric_{ij}$ and
a positive conformal factor $\CF$,
\begin{equation}\label{eq:CMetric}
\SMetric_{ij}=\CF^4\CMetric_{ij},
\end{equation}
and the extrinsic curvature is
split into trace and trace-free parts,
\begin{equation}\label{eq:K-split}
\ExCurv_{ij}=\A_{ij}+\frac{1}{3}\SMetric_{ij}\TrExCurv.
\end{equation}
The freely specifiable data consist of the conformal metric $\CMetric_{ij}$, 
its
time derivative $\dtCMetric_{ij}\equiv\partial_t\CMetric_{ij}$ (which
is taken to be trace free), the mean curvature
$\TrExCurv\equiv K_{ij}\SMetric^{ij}$, and its time derivative
$\partial_tK$.  It follows that the trace-free part of the extrinsic
curvature takes the form
\begin{equation}\label{eq:A}
\A_{ij}=\frac{1}{2\Lapse}\left[\SLong{\Shift}_{ij}-\CF^4\dtCMetric_{ij}\right]=
\CF^{-2}\CA_{ij},\quad \CA_{ij}=\frac{1}{2\CLapse}\left[\CLong{\Shift}_{ij}-\dtCMetric_{ij}\right],
\end{equation}
where
\begin{equation}
\SLong{\Shift}^{ij}\equiv
2\SCD^{(i}\Shift^{j)}-\frac{2}{3}\SMetric^{ij}\SCD_k\Shift^k, \quad
\CLong{\Shift}^{ij}\equiv
2\CCD^{(i}\Shift^{j)}-\frac{2}{3}\CMetric^{ij}\CCD_k\Shift^k.
\label{eq:KillingOperatorDefinitions}
\end{equation}
The symbols $\SLong{\Shift}^{ij}$ and $\CLong{\Shift}^{ij}$ represent
the conformal Killing operators in physical and conformal
space, respectively, and are related by
$\SLong{\Shift}^{ij}=\CF^{-4}\CLong{\Shift}^{ij}$.  Indices on
conformal tensors are raised and lowered with the conformal metric,
for example, $\CLong{\Shift}_{ij}\equiv
\CMetric_{ik}\CMetric_{jl}\CLong{\Shift}^{kl}=\CF^{-4}\SLong{\Shift}_{ij}$.
Furthermore, $\SCD$ and $\CCD$ denote the physical and conformal
spatial covariant derivative operators, and the conformal lapse is defined by
$\Lapse=\CF^6\CLapse$.  Inverting Eq.~(\ref{eq:A}) yields
\begin{equation}\label{eq:dtCMetric}
\dtCMetric_{ij}=\partial_t \CMetric_{ij}
=-2\CLapse\CA_{ij}+\CLong{\Shift}_{ij}.
\end{equation}

Substituting these relations into the constraint equations and into
the evolution equation for the extrinsic curvature, one arrives at a
system of five elliptic equations, often referred to as the extended
conformal thin sandwich (XCTS) equations:\numparts
\begin{eqnarray}\label{eq:XCTSa}
\fl \CCDu^2\CF-\frac{1}{8}\CRicciS\CF-\frac{1}{12}\TrExCurv^2\CF^5 
+\frac{1}{8}\CF^{-7}\CA^{ij}\CA_{ij} = 0,\\
\label{eq:XCTSb}
\fl\CCD_j\Big(\frac{1}{2\CLapse}\CLong{\Shift}^{ij}\Big)
-\frac{2}{3}\CF^6\CCDu^i\TrExCurv
-\CCD_j\Big(\frac{1}{2\CLapse}\dtCMetric^{ij}\Big) = 0,\\
\fl\CCDu^2(\CLapse\CF^7)-(\CLapse\CF^7)\bigg[\frac{\CRicciS}{8}\!+\!\frac{5}
{12}\TrExCurv^4\CF^4\!
+\!\frac{7}{8}\CF^{-8}\CA^{ij}\CA_{ij}\bigg]
\label{eq:Lapse2}
=-\CF^5(\dtime\TrExCurv-\Shift^k\partial_k\TrExCurv).
\label{eq:XCTSc}
\end{eqnarray}
\endnumparts 
Here $\CRicci$ denotes the trace of the Ricci tensor of $\CMetric_{ij}$.
These equations are to be solved for $\CF$, $\CLapse$ and $\Shift^i$;
given a solution, the physical initial data $(\SMetric_{ij}, \ExCurv_{ij})$ are
obtained from Eqs.~(\ref{eq:CMetric})--(\ref{eq:A}).

Note that a solution of the XCTS equations includes a shift vector
$\Shift^i$ and a lapse function $\Lapse=\CF^6\CLapse$.  If these
values of lapse and shift are used in an evolution of the
constructed initial data, then the time derivative of the mean
curvature will initially equal the freely specifiable quantity
$\partial_t\TrExCurv$, and the trace-free part of the time derivative
of the metric will initially equal $\CF^4\dtCMetric_{ij}$. Thus, the
free data of the XCTS equations allow direct control of certain
time derivatives in the evolution of the initial data.

The next step is to choose the free data that correspond to the desired
physical configuration.  The quasi-equilibrium quasi-circular orbit
method of constructing binary black holes~\cite{Cook2004,Caudill-etal:2006} 
(see also Refs.~\cite{Gourgoulhon2001,Grandclement2001,Cook2002}) 
provides a framework
for many of these choices.  This method is based on the fact that the
inspiral time scale for a binary compact object is much larger than
the orbital time scale, so that time derivatives should be very small
in the co-rotating coordinate system.  Furthermore, the black holes
should be in equilibrium, which provides conditions on the expansion
$\NullExpansion$ and shear $\NullShear_{ij}$ of the outgoing
null geodesics passing through the horizon.  The
complete set of physically motivated choices for the free data within the
quasi-equilibrium method are \numparts
\begin{eqnarray}
\dtCMetric_{ij}=0,\label{eq:utilde=0}\\
\partial_t\TrExCurv=0,\label{eq:dtK=0}\\
\CF\to 1,\; \Lapse\to 1,\;  \quad\mbox{as }r\to\infty,\label{eq:AsympFlat}\\
\Shift^i\to (\mathbf{\OmegaOrbitID}\times\mathbf{r})^i, \quad\mbox{as }r\to\infty,\label{eq:OuterBC}\\
\partial_t \mbox{ is tangent to ${\cal S}_{\rm AH}$, }\label{eq:dtAH=0}\\
\NullExpansion=0\quad \mbox{on ${\cal S}$},\label{eq:AH}\\
\NullShear_{ij}=0\quad \mbox{on ${\cal S}$},\label{eq:ZeroShear}
\end{eqnarray}
\endnumparts where ${\cal S}$ denotes the location of
the apparent horizons in the initial data surface, and
${\cal S}_{AH}$ is the world tube of the apparent
horizon obtained by evolving the initial data with lapse $\Lapse$ and
shift $\Shift^i$.  The first two
conditions are the assumptions that the time derivatives are small.
The boundary conditions in Eqs.~(\ref{eq:AsympFlat}) and~(\ref{eq:OuterBC})
enforce asymptotic flatness and co-rotation.  The orbital frequency
$\OmegaOrbitID$ entering Eq.~(\ref{eq:OuterBC}) can be chosen by the
effective potential method~\cite{Cook1994} or the Komar-mass
ansatz~\cite{Gourgoulhon2001}, with similar
results~\cite{Caudill-etal:2006}.

To discuss the remaining conditions, we need to introduce
a few additional geometrical quantities.  Denote by $\SSpatialNormal^i$ and
$\CSpatialNormal^i$ the physical and conformal outward-pointing spatial
unit normals to ${\cal S}$.  They obey the
relations
\begin{equation}\label{eq:ConfScalings}
\SSpatialNormal^i\SSpatialNormal^j\SMetric_{ij}=1,\quad\CSpatialNormal^i\CSpatialNormal^j\CMetric_{ij}=1,\quad\SSpatialNormal^i=\CF^{-2}\CSpatialNormal^i.
\end{equation}
Then introduce the induced metric on ${\cal S}$ in physical and
conformal space by
$\STwoMetric_{ij}=\SMetric_{ij}-\SSpatialNormal_i\SSpatialNormal_j$,
and
$\CTwoMetric_{ij}=\CMetric_{ij}-\CSpatialNormal_i\CSpatialNormal_j$,
respectively.  Because $n_\mu\SSpatialNormal^\mu$=0, the space-time
components of the unit normal are given by $\SSpatialNormal^\mu=[0,
  \SSpatialNormal^i]$.  The outward-pointing null normal to ${\cal S}$
can then be written as
\begin{equation}\label{eq:NullNormal}
\NullNormal^\mu=\frac{1}{\sqrt{2}}\left(n^\mu+\SSpatialNormal^\mu\right).
\end{equation}

Equation (\ref{eq:dtAH=0}) simply means that the apparent horizon is
initially at rest when the initial data is evolved in the co-rotating
coordinate system.  It implies that the shift must take the form
\begin{equation}\label{eq:ShiftBC}
\Shift^i=\Lapse\SSpatialNormal^i+\Shift_{||}^i \quad\mbox{on ${\cal S}$},
\end{equation}
where $\Shift_{||}^i$ is tangent to ${\cal S}$.  Equation (\ref{eq:AH}) ensures that ${\cal S}$ is an apparent horizon, and implies a boundary
condition on the conformal factor, 
\begin{equation}\label{eq:AH-BC}
\CSpatialNormal^k\partial_k \CF = 
-\frac{\CF^{-3}}{8\CLapse}\CSpatialNormal^i\CSpatialNormal^j
\left[\CLong{\Shift}_{ij}-\dtCMetric_{ij}\right]
-\frac{\CF}{4}\,\CTwoMetric^{ij}\CCD_i\CSpatialNormal_j
+\frac{1}{6}\TrExCurv\CF^3.
\end{equation}
Finally,
Eq.~(\ref{eq:ZeroShear})---which forces the apparent horizon to be in
equilibrium---restricts $\Shift_{||}^i$ to be a conformal Killing vector
within the surface ${\cal S}$,
\begin{equation}\label{eq:Skilling1}
\CLong{_{\cal S}\Shift_{||}}^{ij}\equiv 2 \tilde{D}^{(i}\Shift_{||}^{j)}-\CTwoMetric^{ij}\tilde{D}_k\Shift_{||}^k=0,
\end{equation}
where $\tilde{D}_i$ is the covariant derivative compatible with
$\CTwoMetric_{ij}$.  As discussed in detail
in Refs.~\cite{Cook2004,Caudill-etal:2006}, $\Shift_{||}^i$ controls the
spin of the black holes {\em in addition} to the spin required for
co-rotation.

Quasi-equilibrium considerations have now led us to choices for half
of the free data ($\dtCMetric_{ij}$ and
$\partial_t\TrExCurv$) 
for the XCTS equations, and for all boundary conditions except a lapse
boundary condition on the horizon ${\cal S}$.  As argued
in Ref.~\cite{Cook2004}, Eqs.~(\ref{eq:utilde=0})--(\ref{eq:dtAH=0}) are
compatible with any spin of the black holes, with any choice of
boundary conditions for the lapse on ${\cal S}$, and with
any choice of $\CMetric_{ij}$ and $\TrExCurv$.
For concreteness, we choose
\numparts
\begin{eqnarray}
\CMetric_{ij}=\SFlatMetric_{ij},\label{eq:ConfFlatness}\\
\TrExCurv=0,\label{eq:K=0}\\
\partial_r(\Lapse\CF)=0\quad\mbox{on }{\cal S}\label{eq:LapseBC},
\end{eqnarray} 
\endnumparts
where $\SFlatMetric_{ij}$ is the Euclidean metric.  The last two conditions,
Eqs.~(\ref{eq:K=0}) and (\ref{eq:LapseBC}), are
gauge choices~\cite{Cook2004}.  The choice of the conformal metric, 
however, does influence the physical gravitational radiation degrees of 
freedom of the system.  Since a black hole binary is not conformally flat 
at second post-Newtonian order~\cite{Rieth:1997}, our simple choice of 
conformal flatness, Eq.~(\ref{eq:ConfFlatness}), is probably responsible for
the initial burst of unphysical gravitational radiation
found in the evolution of these initial data.

\subsection{Initial data in an asymptotically inertial frame}
\label{sec:QC-inertial}

It is possible to re-formulate the quasi-equilibrium method in
asymptotically inertial coordinates in such a way that {\em identical}
physical initial data are obtained.  To do so, we solve the XCTS
Eqs. (\ref{eq:XCTSa})--(\ref{eq:XCTSc}) with the same choices for
the free data and boundary conditions, except that Eqs.
(\ref{eq:OuterBC}) and (\ref{eq:dtAH=0}) are replaced by
\numparts
\begin{eqnarray}\label{eq:QC-inertial1}
\Shift^i\to 0\;\;\mbox{as }r\to\infty,\\
\label{eq:QC-inertial2}
\partial_{t}+\xirot^i\partial_i\;\; \mbox{ is tangent to ${\cal S}_{\rm AH}$,
where }\xirot^i=(\mathbf{\OmegaOrbitID}\times\mathbf{r})^i.
\end{eqnarray}\endnumparts
The second condition implies that the apparent horizons
move initially with velocity $\xirot^i$, i.e.  tangent to circular
orbit trajectories.

Let $(\CF_{\rm co}, \Shift^i_{\rm co}, \Lapse_{\rm co})$ be the solution to the 
XCTS equations in
the co-rotating coordinates.  We show in~\ref{app:A} that the
solution in the asymptotically inertial coordinates is
$(\CF,\Shift^i,\Lapse)=(\CF_{\rm co}, \Shift^i_{\rm co}-\xirot^i,\Lapse_{\rm co})$,
and that this solution leads to the same physical metric
$\SMetric_{ij}$ and extrinsic curvature $\ExCurv_{ij}$ as the original
solution in co-rotating coordinates.  The proof of this relies on two
observations: First, the shift enters the XCTS equations and the boundary
conditions (almost) solely through the conformal Killing operator,
${\CLong\Shift}^{ij}$; and second, $\xirot^i$ is a
conformal Killing vector, $\CLong{\xirot}^{ij}=0$, for the
conformally flat case considered here.  Hence the term 
$-\xirot^i$ that is added to $\Shift^i_{\rm co}$ drops out of the equations.

In~\ref{app:A}, we also show that Eq.~(\ref{eq:QC-inertial2}) and 
the shear condition Eq.~(\ref{eq:ZeroShear}) require the shift
on the inner boundary ${\cal S}$ to take the form
\begin{equation}\label{eq:ShiftBC-inertial}
\Shift^i=\Lapse\SSpatialNormal^i-\xirot^i+\zeta^i\quad\mbox{on ${\cal S}$},
\end{equation}
where $\zeta^i$ is a vector that must be tangent to ${\cal S}$ 
($\zeta^i\SSpatialNormal_i=0$) and must
be a conformal Killing vector within the surface ${\cal S}$: 
\begin{equation}\label{eq:Lzeta=0}
\NullShear_{ij}=0\quad\Leftrightarrow\quad
 0=\CLong{_{\cal S}\zeta}^{ij}.
\end{equation}

Comparing
Eq.~(\ref{eq:ShiftBC-inertial}) with Eq.~(\ref{eq:ShiftBC}), we see that
the vector $\zeta^i$ plays the role of $\Shift_{||}^i$ in the earlier
treatment; choosing it as a rotation within
${\cal S}$ will impart additional spin to the black holes in addition
to co-rotation, as described in detail in Ref.~\cite{Caudill-etal:2006}.
Note that at large
radii the comoving shift $\Shift^i_{\rm co}$ is a 
pure rotation, since
\(\Shift^i_{\rm co} \rightarrow \xirot^i\) [Eq.~(\ref{eq:QC-inertial1})] and
\(\nabla^j \xirot^i\) is antisymmetric [Eq.~(\ref{eq:QC-inertial2})].

\subsection{Initial data with nonzero radial velocity}
\label{sec:inspiral}

After rewriting the standard quasi-equilibrium method in an asymptotically
inertial frame,
it is straightforward to incorporate nonzero initial radial velocities 
for the black holes.  As
discussed in Sec.~\ref{sec:QC-inertial}, quasi-circular initial data can be generated by specifying
that the horizons move initially on circles in an asymptotically inertial
coordinate system.   This is accomplished by the shift boundary
conditions in Eqs.~(\ref{eq:QC-inertial1}) and~(\ref{eq:QC-inertial2}).
We include initial radial velocities simply by requiring the black holes 
to move initially on inspiral rather than
circular trajectories. 

Consider the problem of giving a black hole located a distance $r_0$ from the
origin an initial radial velocity $v_r$.  This can easily be
accomplished by replacing the boundary conditions
in Eqs.~(\ref{eq:QC-inertial1}) and (\ref{eq:QC-inertial2}) with \numparts
\begin{eqnarray}\label{eq:QC-inspiral1}
\beta^i\to 0\;\; \mbox{as $r\to\infty$},\\
\label{eq:QC-inspiral2} \partial_t+\xiinsp^i\partial_i\; \mbox{
  is tangent to ${\cal S}_{\rm AH}$, where }
\xiinsp^i\equiv(\mathbf{\OmegaOrbitID}\times
\mathbf{r})^i+v_r \frac{r^i}{r_0}.
\end{eqnarray}
\endnumparts 
As before,
we place the center of rotation at the origin of the coordinate
system.  
Note that $\xiinsp^i$ is still a conformal Killing
vector, $\CLong{\xiinsp}^{ij}=0$, for the conformally flat case considered
here.
Therefore the analysis in~\ref{app:A}
of the boundary
conditions in Eqs.~(\ref{eq:QC-inertial2}) and (\ref{eq:ZeroShear}) 
also applies to Eqs.~(\ref{eq:QC-inspiral2}) and
(\ref{eq:ZeroShear}),
and so we find that the inner shift boundary condition must be of the form
\begin{equation}\label{eq:ShiftBC-inspiral}
\Shift^i=\Lapse\SSpatialNormal^i-\xiinsp^i+\zeta^i,\quad\mbox{on ${\cal S}$},
\end{equation}
where $\zeta^i$ is a conformal Killing vector within $\cal S$. 

The boundary conditions in Eqs.~(\ref{eq:QC-inspiral1}) 
and (\ref{eq:QC-inspiral2})
depend on two parameters, the orbital frequency $\OmegaOrbitID$ and a
radial velocity $v_r$ (or, more precisely, an overall expansion
factor $v_r/r_0$, reminiscent of the Hubble constant).  For unequal
mass binary systems the needed radial velocities for each hole would
be different, but the needed expansion factors, $v_r/r_0$, are expected to 
be the same
for the two holes.

The changes discussed in Sec.~\ref{sec:QC-inertial} are superficially
similar to the changes discussed in Sec.~\ref{sec:inspiral}, yet
the former amounts to a mere coordinate transformation while the
latter produces different physical initial data.  This can be
understood by noting that the change from co-rotating coordinates
[Eqs.~(\ref{eq:OuterBC}) and (\ref{eq:dtAH=0})] to inertial
coordinates [Eqs.~(\ref{eq:QC-inertial1}) and (\ref{eq:QC-inertial2})]
is accomplished by adding the {\em same} conformal Killing vector
field $\xirot^i$ to the shift
at both inner and outer boundaries, but the change
from Eqs.~(\ref{eq:OuterBC}) and
  (\ref{eq:dtAH=0}) to initial data with nonzero radial velocity
[Eqs.~(\ref{eq:QC-inspiral1}) and (\ref{eq:QC-inspiral2})] is
accomplished by adding {\em different} conformal Killing fields to the
shift on different boundaries: $\xirot^i$ at the outer
boundary and $\xiinsp^i$ at the inner boundaries. 
Only in the former case can the change be expressed
as a global transformation of the shift of the
form $\Shift^i\to\Shift^i+\xirot^i$.


\section{Numerical methods}
\label{sec:numerical-methods}

The initial value equations are solved with the pseudo-spectral
elliptic solver described in Ref.~\cite{Pfeiffer2003}.  This elliptic solver
has been updated to share the more advanced infrastructure of our
evolution code and is now capable of handling cylindrical subdomains.
This increases its efficiency by about a factor of three over the results
described in Ref.~\cite{Pfeiffer2003} for binary black hole initial
data.

The Einstein evolution equations are solved with the
pseudo-spectral evolution code
described in Ref.~\cite{Scheel2006}.  This code evolves a first-order
representation~\cite{Lindblom2006} of the generalized harmonic
system~\cite{Friedrich1985,Pretorius2005c}.  We use boundary 
conditions~\cite{Lindblom2006} designed to
prevent the influx of unphysical constraint violations and undesired
incoming gravitational radiation, while allowing the outgoing
gravitational radiation to pass freely through the boundary.  The code
uses a fairly complicated domain decomposition.  Each black hole is
surrounded by three concentric spherical shells, with the inner
boundary of the inner shell just inside the horizon.  The inner shells
overlap a structure of 24 touching cylinders, which in turn overlap
a set of outer spherical shells---centered at the origin---which
extend to large outer radius.  Outer boundary conditions are imposed 
only on the outer surface of the largest outer spherical shell.  
We vary the location of the outer
boundary by adding more shells at the outer edge.  Since all
outer shells have the same angular resolution, the cost of placing the
outer boundary farther away (at full resolution) increases only
linearly with the radius of the boundary.  Some of the details of the domain
decompositions used for the simulations presented here are given in Table
\ref{tab:evolutions}.

\section{Choice of orbital frequency and radial velocity}
\label{sec:3}

We now describe how to construct binary black hole initial data sets
with low orbital eccentricity.
This is done by tuning the freely adjustable orbital parameters
$\Omega_0$ and $v_r$ iteratively to reduce the
eccentricity of the inspiral trajectories. 
For each iteration we choose trial orbital parameters $\Omega_o$ and
$v_r$, evolve the corresponding initial data,
analyze the resulting trajectories of the black holes, 
and update the orbital parameters to reduce any oscillatory 
behavior in quantities like the
coordinate separation of the black holes $d(t)$, the proper separation between
the horizons $s(t)$, or the orbital frequency $\omega(t)$. 
All of these quantities (and many others) exhibit similar oscillatory
behavior; we choose $d(t)$ as our primary diagnostic during the tuning
process because it is most easily accessible during the
evolutions.  

\begin{figure}
\centerline{\includegraphics[width=2.7in]{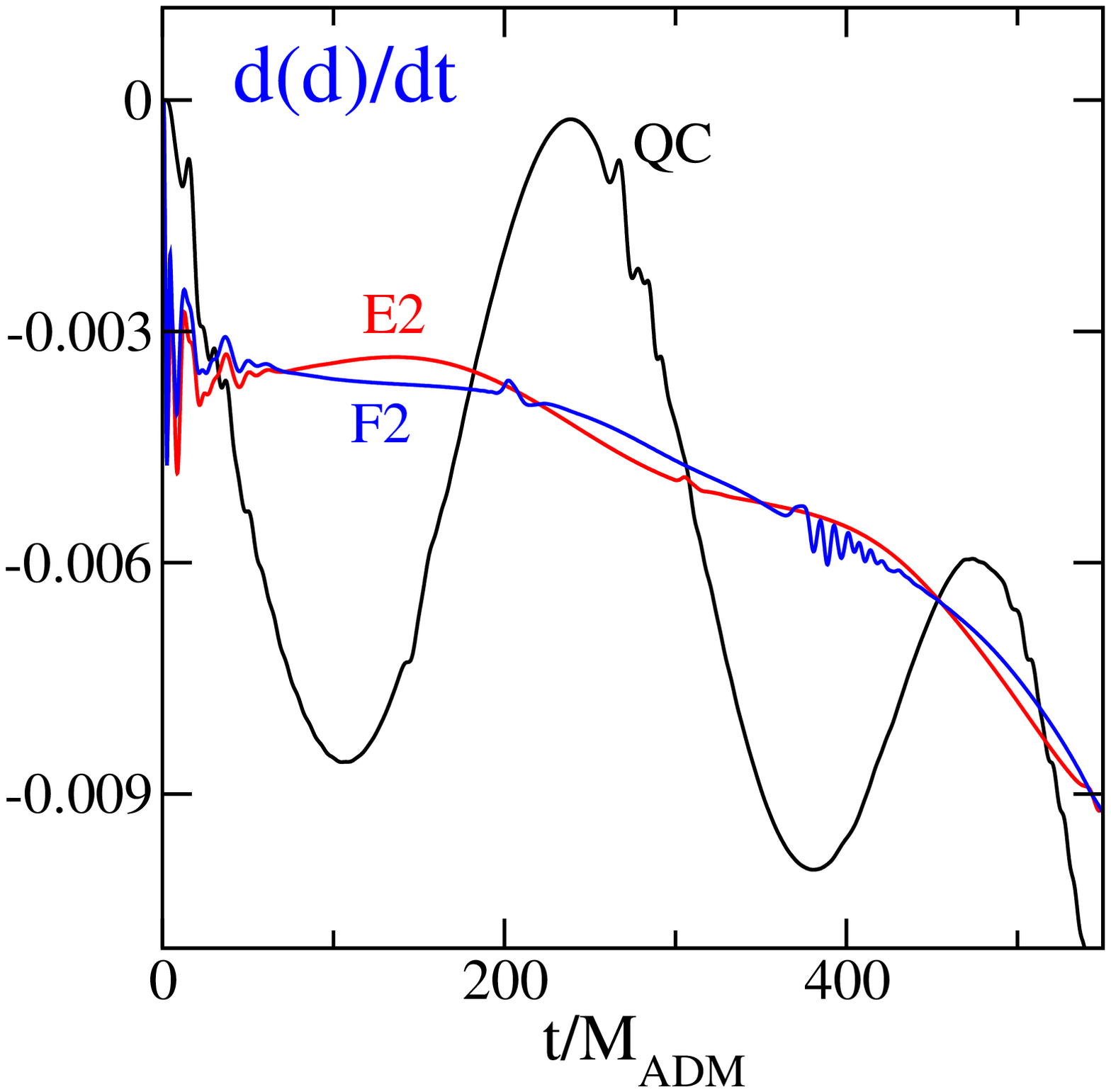}
$\qquad
$\includegraphics[width=2.7in]{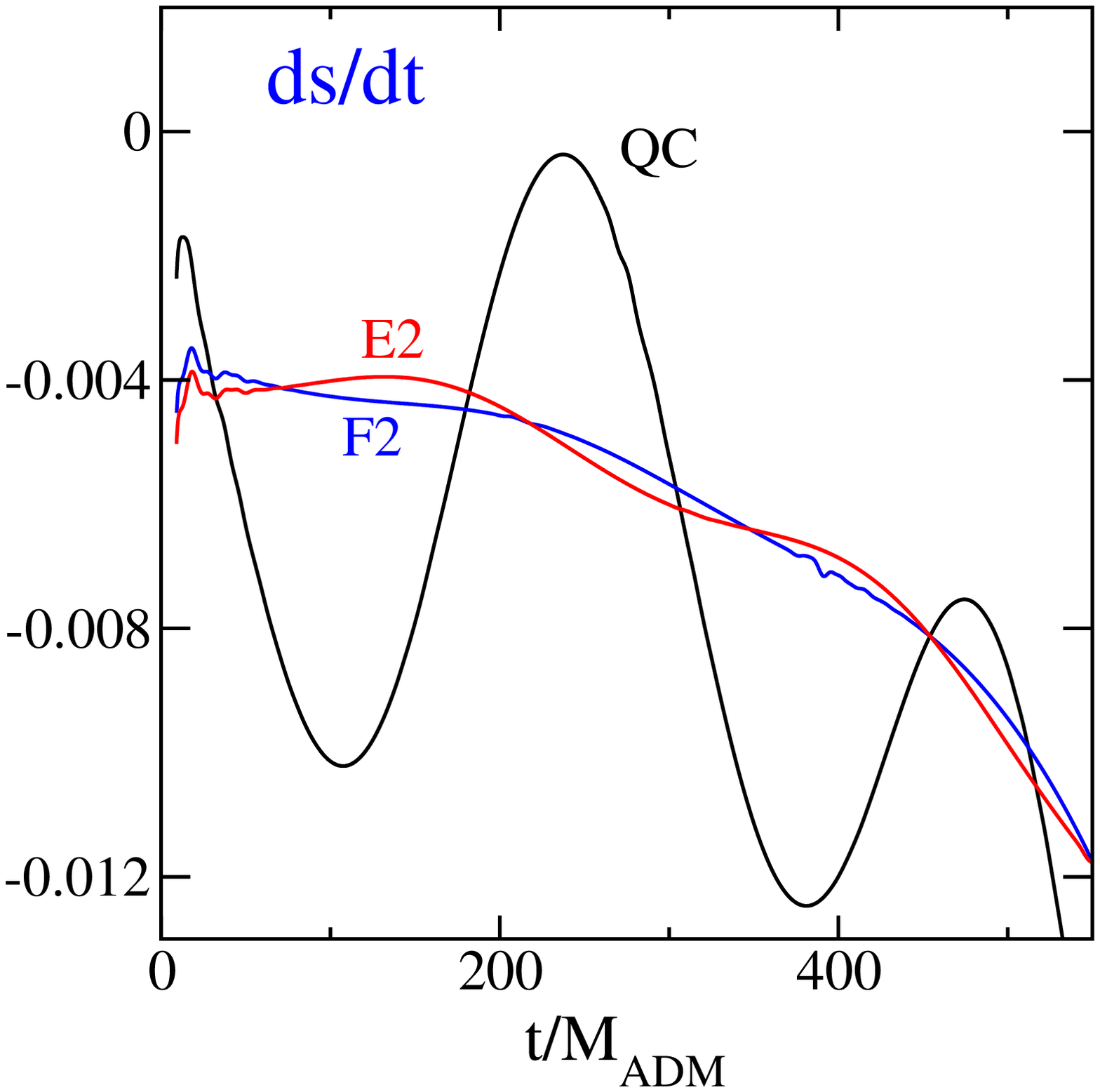}
}
\caption{\label{fig:RadialVelocity} Radial velocity during
  evolutions of quasi-circular and low-eccentricity initial data.  The
  left panel shows the coordinate velocity $\dot{d}(t)$, the right
  panel the velocity determined from the intra-horizon proper separation 
  $\dot{s}(t)$.  }
\end{figure}

To make this procedure quite explicit, we begin by evolving quasi-circular
initial data for about two orbits.  Then we measure the time derivative of 
the measured coordinate separation of the holes $\dot d(t)$
(in the asymptotic inertial coordinates used in our code~\cite{Scheel2006})
as illustrated for example in Fig.~\ref{fig:RadialVelocity}.  We
fit this measured $\dot d(t)$ to a function of the form:
\begin{equation}\label{eq:fit}
\dot d(t) = A_0 + A_1 t + B \sin(\omega t + \varphi),
\end{equation}
where $A_0$, $A_1$, $B$, $\omega$, and $\varphi$ are constants determined
by the fit.  The $A_0+A_1 t$ part of the solution represents the smooth
inspiral, while the  $ B\sin(\omega t + \varphi)$ part represents
the unwanted oscillations due to the eccentricity of the orbit.
For a nearly circular Newtonian orbit, $B$ is related to the eccentricity
$e$ of the orbit by $e=B/\omega d$.  So reducing the orbital eccentricity
is equivalent to reducing $B$.
The values of the orbital parameters
$\Omega_0$ and $v_r$ are now adjusted iteratively to make
the coefficient $B$ in this fit as small as desired.  After each
adjustment of $\Omega_0$ and $v_r$, the initial value equations
described in Sec.~\ref{sec:2} [in particular, using the boundary
condition~(\ref{eq:QC-inspiral2}) which depends on
$\Omega_0$ and $v_r$] are solved completely (to the 
level of numerical truncation error). 

For this paper, our goal is to reduce $B$,
and hence the orbital eccentricity, by about
a factor of ten compared to quasi-circular initial data.  This level
of reduction is sufficient to
allow us to evaluate the significance of
the orbital eccentricity inherent in quasi-circular initial data.
A variety of methods could be used to find orbital paramters that make
$B$ small.  One possibility is simply to evaluate $B(\Omega_0,v_r)$
numerically as described above, and then to use standard numerical
methods to solve the equation $B(\Omega_0,v_r)=0$.  Since
our goal in this paper is to reduce $B$ by about a factor of ten,
simple bisection root finding methods are sufficient.  

A more efficient method is to use our knowledge of the behavior of
nearly circular orbits to make informed estimates of the needed
adjustments in the orbital parameters.  Evaluating the fit Eq.~(\ref{eq:fit}) 
at the initial time $t=0$, we see that the ellipticity-related component 
$B\sin(\omega t+\varphi)$ contributes $B\sin(\varphi)/2$ to the radial 
velocity of each hole and $B\omega \cos(\varphi)/2$ to its radial acceleration.
(The factor $1/2$ arises because $d$ measures the distance between the holes.)
For a Newtonian binary, this eccentricity-induced
radial velocity can be completely removed
by changing the initial radial velocity by
\begin{equation}\label{eq:vr-update}
\delta v_r = - \frac{B \sin(\varphi)}{2}.
\label{eq:vadjust}
\end{equation}
Furthermore, changing the orbital frequency $\Omega_0$ by a small amount $\delta \Omega_0$
changes the radial acceleration of each black hole by the amount 
$\Omega_0\delta \Omega_0 d_0$, where $d_0=d(0)$ is the initial
separation of the holes.  Thus the change $\delta \Omega_0$ needed
to remove the eccentricity-induced initial radial acceleration,
$B\omega \cos(\varphi)/2$, is
\begin{equation}\label{eq:Omega0-update}
\delta\Omega_0 = -\frac{B\omega\cos(\varphi)}{2 d_0\Omega_0}
\approx-\frac{B\cos(\varphi)}{2 d_0}.
\label{eq:omegaadjust}
\end{equation}
Equations~(\ref{eq:vr-update}) 
and (\ref{eq:Omega0-update}) still hold approximately
for relativistic binaries.
We have found that simultanously adjusting $v_r$ and $\Omega_0$ by 
Eqs.~(\ref{eq:vadjust}) and (\ref{eq:omegaadjust}) typically reduces $B$ by 
about a factor of ten.

The smallest eccentricity data set produced here (by the simple
bisection method described above)
is labeled `F', and the data
from the next to last iteration of this method is labeled `E'.  These initial
data sets, together with the quasi-circular data labeled `QC'
were evolved with multiple numerical resolutions and with multiple
outer boundary locations; Table~\ref{tab:evolutions} summarizes
these evolutions.  The orbital frequency used in the final
evolution is
only 0.6 per cent larger than the value of $\Omega_0$ used in
the quasi-circular case.  As expected, this change is comparible to 
the magnitude of the radial velocity $v_r$ in the low eccentricity case.
The smallness of these quantities shows that the quasi-circular 
approximation is quite good.

\begin{table}
\caption{\label{tab:evolutions}Summary of evolutions presented in this
  paper. The labels `QC', `E' and `F' refer to the different initial
  data sets, with numerical suffix (`E1', `E2' etc.)  denoting different values of the initial outer boundary radius of the evolutions, $R_{\rm outer}$.
}
\begin{indented}
\item[]

\begin{tabular}{c|l|ccccc}
\br
Label & Initial data & $\frac{R_{\rm outer}}{\Eadm}$ & \# outer  &\multicolumn{3}{c}{approx \# of points}\\
&  & & shells & low & med. & high\\
\mr
QC & $\Eadm\Omega_0\!=\!0.029792,\;v_r\!=\!0.0$ & $133$ & 8 & $ 52^3$ & $ 64^3$ & $ 76^3$ \\
   & $J_{\rm ADM}/\Eadm^2=0.98549$ & \\
& $M_{\rm irr}/\Eadm=0.50535$ \\\hline

%
E1 & $\Eadm\Omega_0\!=\!0.029961,\;v_r\!=\!-0.0017$ & $171$ & 10 & $ 59^3$ & $ 66^3$ & $ 74^3$ \\
%
E2 & $J_{\rm ADM}/\Eadm^2=0.99172$  & $293$ & 18 & $ 64^3$ & $ 72^3$ & $ 81^3$ \\
   & $M_{\rm irr}/\Eadm=0.50524$ \\\hline
%
F1 & $\Eadm\Omega_0\!=\!0.029963,\;v_r\!=\!-0.0015$ & $133$ & 8 & $ 52^3$ & $ 64^3$ & $ 76^3$  \\
%
F2 &  $J_{\rm ADM}/\Eadm^2=0.99164$ & $190$ & 12 & $ 55^3$ & $ 66^3$ & $ 78^3$ \\
%
F3 & $M_{\rm irr}/\Eadm=0.50525$  & $419$ & 28 & $ 62^3$ & $ 74^3$ & $ 87^3$ \\
\br
\end{tabular}
\end{indented}
\end{table}

\begin{figure}
\centerline{\includegraphics[width=4.4in]{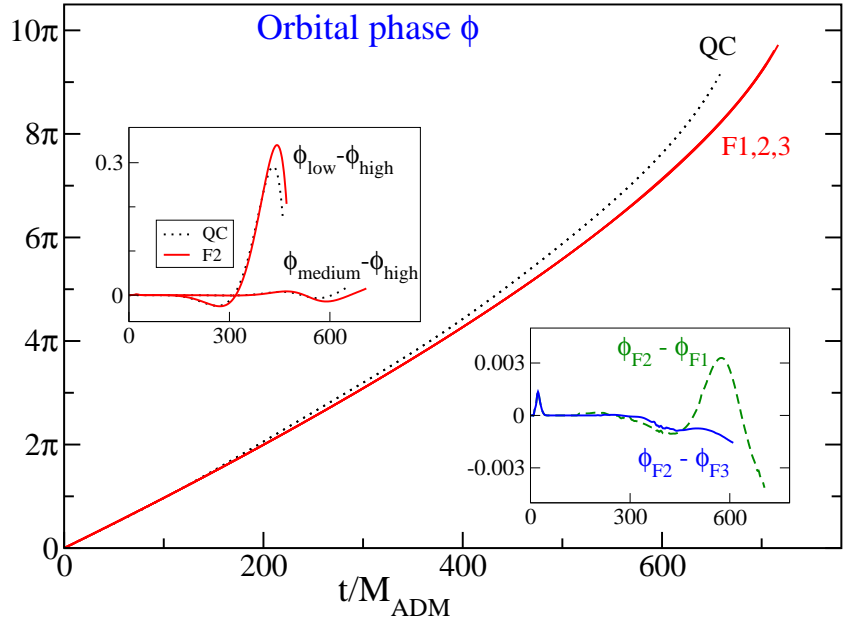}
}
\caption{\label{fig:OrbitalPhase}Evolution of the orbital phase.  
  The main panel
  shows the phase of the trajectories of the centers of the apparent
  horizons as a function of time for the quasi-circular (dotted curves) and
  low-eccentricity (solid curves) initial data.  The top left inset shows
  the phase differences between different resolution runs,
  which decreases at higher resolutions.
  The lower right inset
  shows the difference in the orbital phase between evolutions with
  different outer boundary locations.  }
\end{figure}


\Fref{fig:OrbitalPhase}
shows the orbital phase (as measured by the coordinate locations of
the centers of the apparent horizons) for the evolutions of
quasi-circular initial data, QC, and the least-eccentric
initial data, F1, F2, and F3. (The numerical suffix, F1, F2, etc.,
denotes simulations with different values of the outer boundary radius
as defined in Table~\ref{tab:evolutions}.) 
These evolutions proceed for
about five orbits and then crash shortly before the black holes merge.  The
upper left inset shows differences between the orbital phase computed
with different resolutions for the QC and the F2 runs.  The phase difference
between the high and low resolution runs is $\lesssim 0.35$ radians,
which is a good estimate of the error in the low resolution run. The
phase difference between the medium and high resolution runs
drops to $\approx 0.02$ radians, which can be taken as the error in the medium
resolution run.  Between low and medium resolutions, the error drops
by about a factor of $20$.  Assuming exponential convergence, the
error of the high resolution run should be smaller by yet another
factor of $\sim 20$, i.e. $\lesssim 0.001$ radians.
The lower right inset in Fig.~\ref{fig:OrbitalPhase} 
shows phase differences between evolutions of
the same initial data, but run with different outer boundary radii.  These
differences are small, so we do not expect the influence of the
outer boundary on our results to be significant.  
Our analysis in Sec.~\ref{sec:4} is based mostly on
comparisons between the high resolution QC and F2 runs.

Figure \ref{fig:RadialVelocity} illustrates the radial velocities 
(determined from the time derivatives of both the coordinate and 
the intra-horizon proper separations) for the
quasi-circular run QC and for the two low-eccentricity runs
E and F.  Orbital eccentricity causes periodic oscillations in these
curves; the amplitudes of these oscillations are clearly much smaller in
runs E and F than in run QC. By fitting the proper separation speed $ds/dt$ to
a linear function plus sinusoid, $ds/dt=A_0+A_1 t + B 
\sin(\omega t+\varphi)$, the approximate amplitude of
the oscillations can be estimated.  
We find $B_{\rm QC}\approx 5.5\cdot 10^{-3}$,
$B_{\rm E}\approx 5.8\cdot 10^{-4}$, and $B_{\rm F}\approx 4.1\cdot
10^{-4}$.   This confirms that we have succeeded in our goal of reducing the
oscillations by an order of magnitude.  These fits are not very
accurate because the fit must cover at least one period of the
oscillations, and significant orbital evolution
occurs during this time.  
If we vary the fit interval $40<t/\Eadm<T$ by choosing $T$ between $300$ and
$450$, the quoted amplitudes $A_{\rm QC,E,F}$  change at about the 10\% level.

The coordinate separation $d(d)/dt$ shows some noise at early times as the
binary system equilibrates and an initial burst of `junk' gravitational
radiation travels outward.  There are also short-lived, high-frequency features
apparent in Fig.~\ref{fig:RadialVelocity} at
intermediate times.  The earlier feature occurs at $t/\Eadm\sim 140$
for the QC run, $t/\Eadm\sim 200$ for F2, and 
$t/\Eadm\sim 300$ for E2; these times coincide
with the light-crossing time to the outer boundary. We believe
that this early feature is caused
by a small mismatch between the initial data
and the outer boundary conditions used by the evolution code;
this mismatch produces a
pulse that propagates inward from the
outer boundary starting at $t=0$.
A later (and larger)
feature occurs at $t/\Eadm\sim 280$
for the QC run, $t/\Eadm\sim 400$ for F2, and 
at $t/\Eadm\sim 600$ (off the scale of Fig.~\ref{fig:RadialVelocity}) 
for E2. This later feature
occurs at
twice the light-crossing time, and is caused
by reflection of the
initial `junk' gravitational radiation burst off of the outer boundary.  
The outer boundary conditions used in this paper perform well for the
physical gravitational-wave degrees of freedom~\cite{Lindblom2006},
but comparatively poorly for the gauge degrees of freedom
(as demonstrated in recent tests~\cite{Rinne2007}). These results plus the
observation that the high-frequency features in Fig.~\ref{fig:RadialVelocity}
are greatly
diminished in less gauge-dependent quantities like $ds/dt$ suggests
that these 
features may be caused by perturbations in the gauge or coordinate degrees of
freedom of the system.

\begin{figure}
\centerline{\includegraphics[width=5.4in]{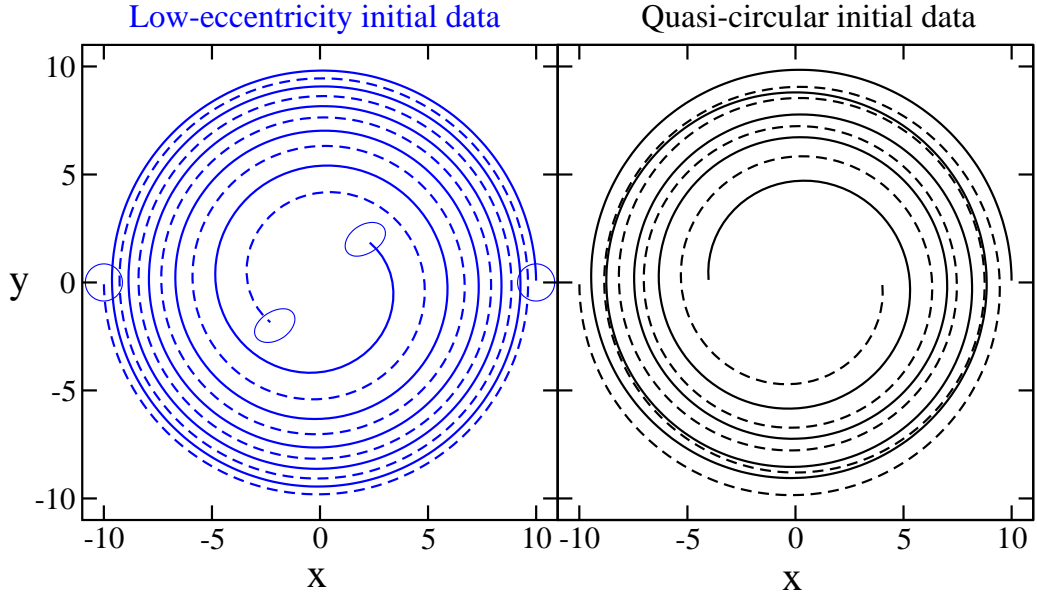}
}
\caption{\label{fig:Trajectories}Trajectories of the center of the
  apparent horizons in asymptotically inertial coordinates for the runs
  E1 (left plot) and QC (right plot).  The solid/dashed line
  distinguish the two black holes; the circles and ellipsoids in the
  left figure denote the location of the apparent horizon at the
  beginning and end of the evolution.}
\end{figure}

Figure \ref{fig:Trajectories} shows the orbital
trajectories of the centers of the black holes during evolutions of
the low-eccentricity initial data E \footnote{We plot the evolution E1
  because it was pushed somewhat closer to merger than the F runs; the
  trajectories of the E runs are indistinguishable from those of the F runs
  on the scale of
  this figure.}, and the quasi-circular initial data QC.  The
low-eccentricity run forms a smooth spiral with no apparent distortion.  In
contrast, the evolution starting from quasi-circular initial data has
clearly visible irregularities.

\section{Comparing quasi-circular and low-eccentricity initial data}
\label{sec:4}

Figures \ref{fig:RadialVelocity} and \ref{fig:Trajectories} show clearly 
that evolutions of the quasi-circular initial data, QC, are not the same as
those of the low-eccentricity initial data, F.  In this section, we
characterize and quantify these differences in more detail.

\subsection{Time shift}
\label{sec:4.1}
\begin{figure}
\centerline{
\includegraphics[scale=0.46]{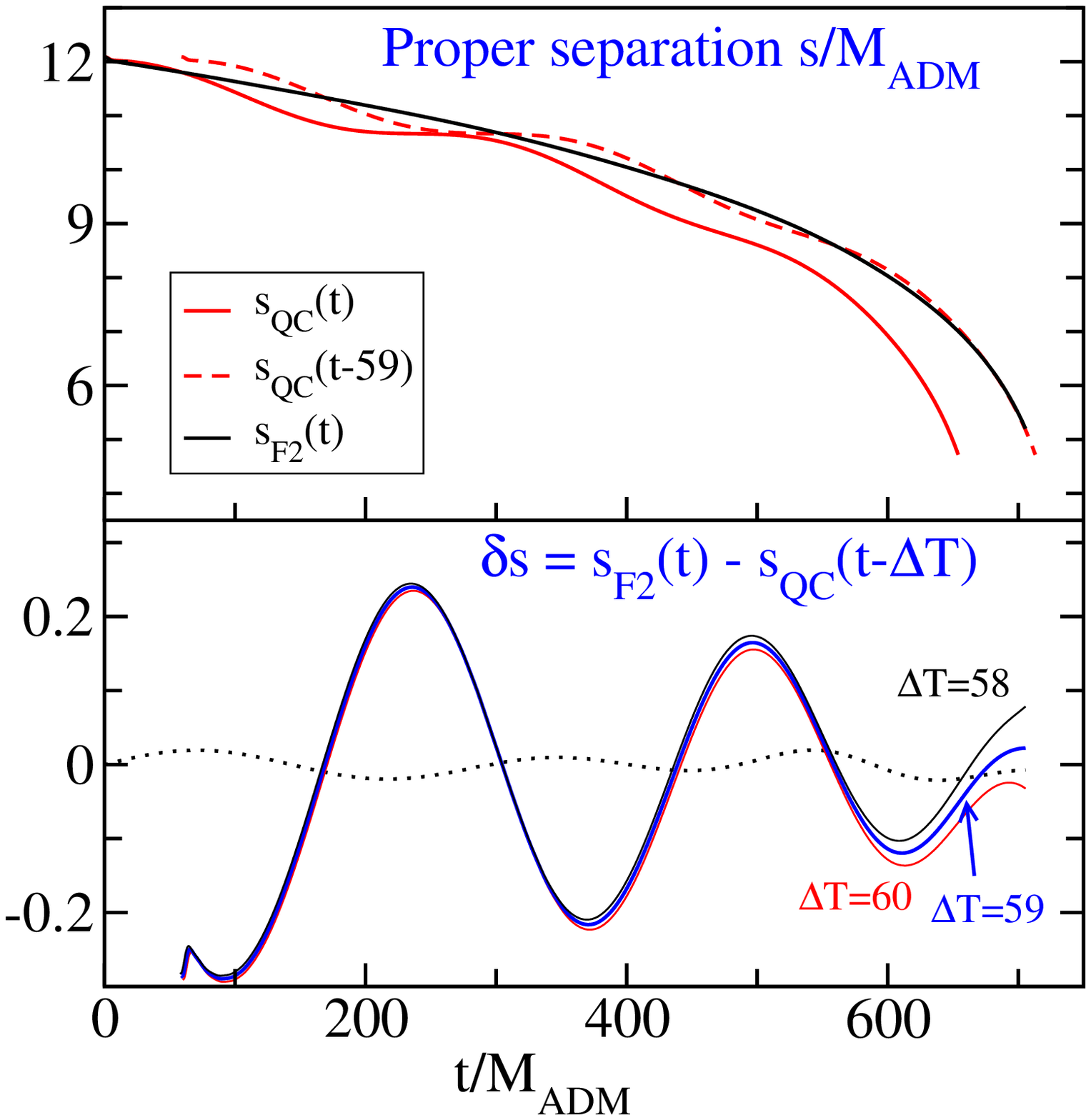}
\;\;\;\includegraphics[scale=0.46]{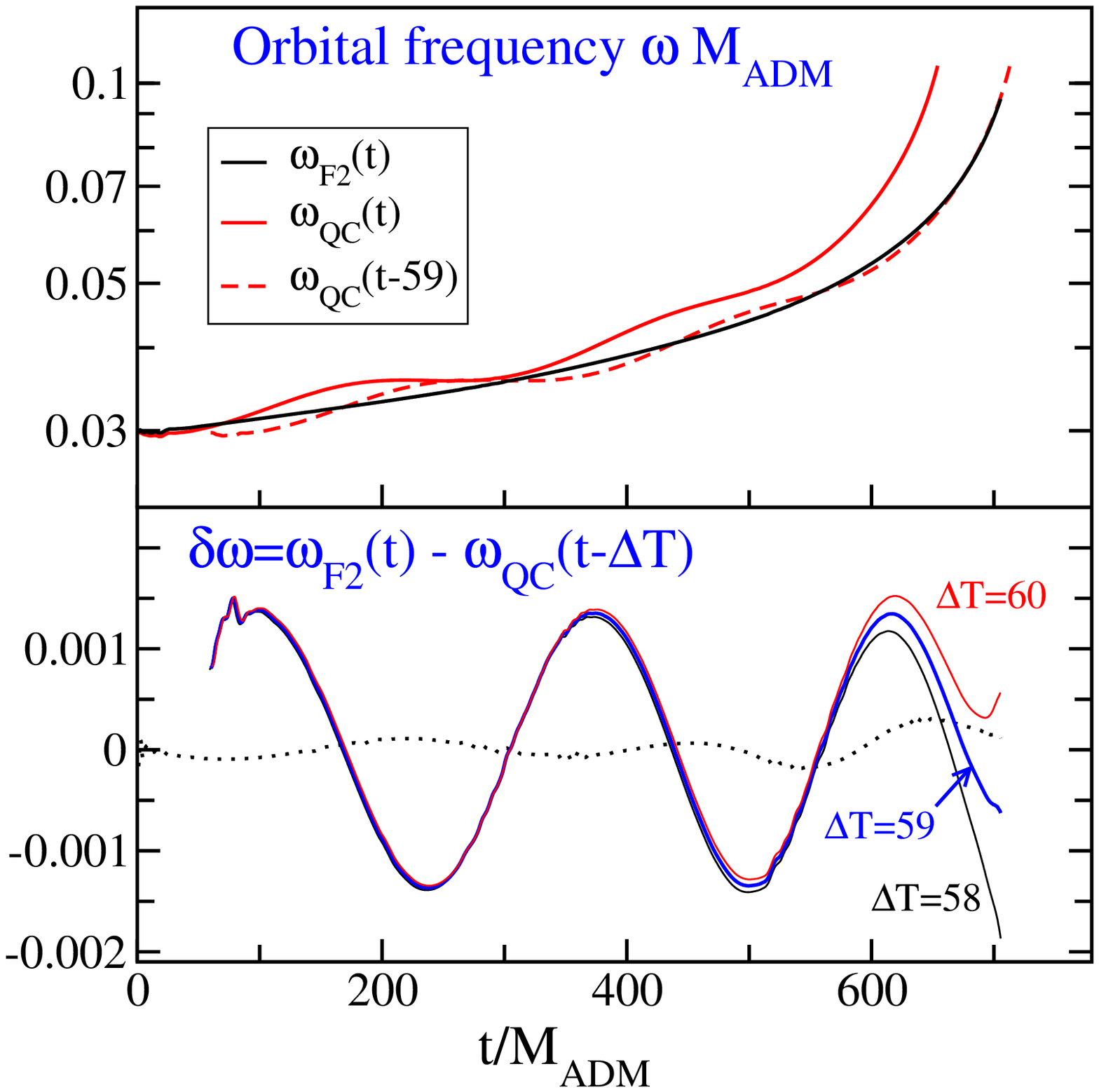}
}
\caption{\label{fig:ProperSep}Proper separation (left) and orbital
  frequency (right) for evolutions of the QC and F initial data.  The
  lower panels show the differences between the time-shifted QC and
  the F2 runs.  
The dotted lines in the lower
  panels show the differences between the E1 and F2 runs, providing an
  estimate of the remaining eccentricity in the F2 run.  }
\end{figure}

The black holes approach each other more quickly in the QC run, 
with the time of coalescence appearing to be about
$60\Eadm$ earlier than in the F2 run.
Figure~\ref{fig:OrbitalPhase}, for example, shows that
the orbital phase increases more quickly
during the QC run, with a late time phase difference of about
$\pi$ (almost a full gravitational wave cycle)
compared to the F2 run.  Similar
differences are also seen in the graphs of 
the proper separation and
orbital frequency shown in the upper panels of
Fig.~\ref{fig:ProperSep}.

We find that most of the difference between the QC and F runs
is just a simple coordinate time shift.
The dashed lines in the upper panels of Fig.~\ref{fig:ProperSep}
represent the QC evolution shifted by $\Delta T=59\Eadm$. With this
time shift, the QC evolution oscillates around the low-eccentricity
F2 run.  
Therefore, the apparent earlier merger time of the QC run is just a
consequence of the fact that coordinate time $t=0$ in the QC run represents
a later stage in the inspiral than it does in the F2 evolution.  
The QC and F2 runs were started with the same spatial coordinate 
separation at $t=0$;
however, this point is the apocenter of the slightly
eccentric QC orbit,  so the point in the F2 run with the same
phase (measured from merger) has
smaller separation.  

The
lower left panel of Fig.~\ref{fig:ProperSep} shows the proper separation
difference,
$\delta\PropSep=\PropSep_{F}(t)-\PropSep_{QC}(t-\Delta T)$, which
emphasizes the oscillations of the QC evolution around the F2 orbit.
These differences are plotted for three different time shifts $\Delta T$.
The right panels of Fig.~\ref{fig:ProperSep} present information
about the orbital angular frequency $\OmegaOrbitEv$ as determined from
the coordinate locations of the centers of the apparent horizons.  The
upper right panel shows $\OmegaOrbitEv$ for evolutions of QC and F2
initial data.  Time-shifting the QC run by the same $\Delta T=59\Eadm$
also lines up the frequency curves very
well.  The lower right plot shows the difference in orbital frequency
between the F2 run and the time-shifted QC run,
$\delta\OmegaOrbitEv=\OmegaOrbitEv_{\rm F}(t)-\OmegaOrbitEv_{\rm
  QC}(t-\Delta T)$.
The differences $\delta\PropSep$ and $\delta\OmegaOrbitEv$ are very
sensitive to the time offset $\Delta T$ applied to the QC run.  In
particular, at late times, when $\PropSep$ and $\OmegaOrbitEv$ vary
rapidly, even a small change in $\Delta T$ causes the differences to
deviate significantly from their expected oscillatory behavior around zero.
Looking at both $\delta\PropSep$ and $\delta\OmegaOrbitEv$, we
estimate a time offset $\Delta T/\Eadm=59\pm 1$ between the QC run and
the F runs.

\subsection{Measuring eccentricity}
\label{sec:4.2}

The evolution of the F initial data appears to have very low orbital
eccentricity, so it can be used as a reference from which
the eccentricity of the QC run can be estimated.  We
can define an eccentricity for the QC evolution, for example, from the 
relative proper separation,
\begin{equation}\label{eq:es}
e_\PropSep=\frac{|\delta\PropSep|}{\PropSep},
\end{equation}
where this equation is to be evaluated at the extrema of
$\delta\PropSep$.  Similarly, we can define a different measure
of eccentricity from the
variations in $\omega_{\rm orbit}$ by evaluating
\begin{equation}\label{eq:eomega}
e_\OmegaOrbitEv=\frac{|\delta\OmegaOrbitEv|}{2\OmegaOrbitEv}
\end{equation}
at the extrema of $\delta\OmegaOrbitEv$.  The factor of two in the
definition of $e_\OmegaOrbitEv$ arises from
angular momentum conservation, which makes the orbital frequency
proportional to the square of the radius of the orbit.  In Newtonian gravity,
$e_\PropSep=e_{\OmegaOrbitEv}$ to first order in eccentricity.  Since
the F initial data results in a factor of ten smaller oscillations in 
$d\PropSep/dt$ than the QC data,
we expect these eccentricity estimates to be affected by the
residual eccentricity of the F run at only the 10\% level.

\begin{figure}
\centerline{\includegraphics[width=3.7in]{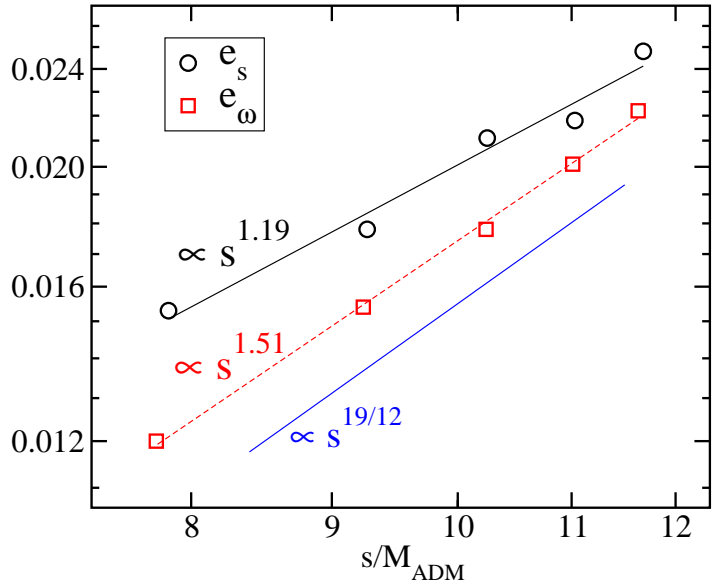}
}
\caption{\label{fig:eccentricity}Orbital eccentricity of the QC evolution
  estimated from variations in proper separation, $e_s$, and from variations
  in orbital frequency, $e_\omega$.  
  Also shown in this log-log plot are best-fit
  power laws to each set of data, as well as the scaling predicted by
  Peters~\cite{Peters1964} with power $19/12\approx 1.58$.

 }
\end{figure}


The orbital eccentricity of the QC run, estimated using 
Eqs.~(\ref{eq:es}) and (\ref{eq:eomega}), is plotted as a function
of proper separation between the black holes in
Fig.~\ref{fig:eccentricity}.  We see that these eccentricities decay 
during the inspiral, as expected.  Within our estimated 10\% errors,
these eccentricities are consistent with a power law dependence on
the proper separation,
$e\propto s^p$.  The eccentricity $e_\PropSep$ based on the
proper separation is consistently somewhat larger than
$e_\OmegaOrbitEv$, and it decays somewhat more slowly.
Peters~\cite{Peters1964} derived the evolution of the orbital
eccentricity during an inspiral due to the emission of gravitational waves 
using the quadrupole approximation.  His result in the $e\ll 1$ limit 
predicts that $e\propto a^{19/12}$, where $a$ is the semi-major
axis of the orbit and where the constant of proportionality depends on the 
initial conditions.  Using $a\approx \PropSep/2$, his formula
predicts that the eccentricity should decay as 
\begin{equation}
e\propto \PropSep^{19/12}.
\end{equation}
Figure~\ref{fig:eccentricity} confirms that $e_\OmegaOrbitEv$
follows this prediction quite closely, while $e_\PropSep$ has
a somewhat smaller power law exponent.

The eccentricities measured here are actually the relative eccentricities of
the QC and the F orbits.  The eccentricity of the QC run that we infer
depends therefore on the residual eccentricity of the F run.  A more
intrinsic approach, used recently by Buonanno et
al.~\cite{Buonanno-Cook-Pretorius:2006}, is to fit some 
eccentricity-dependent quantity to a full cycle (or more) of the orbital data.
This approach yields similar, but somewhat smaller, eccentricities
than those found here (despite our use of a QC orbit having larger
initial separation and so presumably smaller initial eccentricity).

\subsection{Waveform extraction}
\label{sec:4.3}

We now turn our attention to the problem of extracting the gravitational 
wave signals from our numerical simulations using
the Newman-Penrose quantity $\Psi_4$.  Given a spatial
hypersurface with timelike unit normal $n^\mu$, and given a spatial
unit vector $r^\mu$ in the direction of wave propagation, the standard
definition of $\Psi_4$ is the following component of the Weyl
curvature tensor,
\begin{equation}
\Psi_4 = - C_{\alpha\mu\beta\nu} \ell^\mu \ell^\nu \bar{m}^\alpha\bar{m}^\beta,
\label{eq:Psi4Definition}
\end{equation}
where $\ell^\mu \equiv \frac{1}{\sqrt{2}}(n^\mu - r^\mu)$,
and $m^\mu$ is a complex null vector (satisfying $m^\mu \bar{m}_\mu = 1$)
that is orthogonal to $r^\mu$ and $n^\mu$. Here an overbar denotes complex
conjugation.  

For (perturbations of) flat spacetime, $\Psi_4$ is typically evaluated
on coordinate spheres, and in this case the usual
choices for $r^\mu$ and $m^\mu$ are
\numparts
\begin{eqnarray}
  r^\mu &=& \left(\frac{\partial}{\partial r}\right)^\mu,
            \label{eq:FlatspaceRadialTetrad}\\
  m^\mu &=& \frac{1}{\sqrt{2} r}
            \left(\frac{\partial}{\partial \theta} 
            +    i\frac{1}{\sin\theta}\frac{\partial}{\partial \phi}\right)^\mu,
  \label{eq:FlatspaceMTetrad}
\end{eqnarray}
\endnumparts
where $(r,\theta,\phi)$ denote the standard spherical coordinates.
With this choice, $\Psi_4$
can be expanded in terms of spin-weighted spherical harmonics of weight -2:
\begin{equation}
\Psi_4(t,r,\theta,\phi) 
= \sum_{l m} \Psi_4^{l m}(t,r)\, {}_{-2}Y_{l m}(\theta,\phi),
\label{eq:Psi4Ylm}
\end{equation}
where the $\Psi_4^{l m}$ are expansion coefficients defined by this equation.

For curved spacetime, there is considerable freedom in the choice of
the vectors $r^\mu$ and $m^\mu$, and different researchers have made
different choices~\cite{Buonanno-Cook-Pretorius:2006,Fiske2005,%
Beetle2005,Nerozzi2005,%
Burko2006,Campanelli2006,Bruegmann2006}
that are all equivalent in the $r\to\infty$ limit.  
We choose these vectors by first picking an extraction two-surface $\cal E$
that is a coordinate sphere ($r^2=x^2+y^2+z^2$) centered on the 
center of mass of the binary system
(using the global asymptotically Cartesian coordinates
employed in our code).  We choose $r^\mu$ to be the
outward-pointing spatial unit normal to $\cal E$ (that is, we choose
$r_i$ proportional to $\nabla_i r$).  Then we choose $m^\mu$
according to Eq.~(\ref{eq:FlatspaceMTetrad}), using the standard spherical
coordinates $\theta$ and $\phi$ defined on these coordinate spheres.
Finally we use
Eqs.~(\ref{eq:Psi4Definition}) and~(\ref{eq:Psi4Ylm}) to define the
$\Psi_4^{l m}$ coefficients.  Note that our $m^\mu$ is not 
exactly null nor exactly of unit magnitude at finite $r$,
so our definition of $\Psi_4^{l m}$ will disagree with
the waveforms observed at infinity (and with those computed by other
groups).  Our definition does, however,
agree with the standard definition given in
Eqs~(\ref{eq:Psi4Definition})--(\ref{eq:Psi4Ylm}) 
as $r\to\infty$, so our definition only disagrees with the standard
one by a factor of order $1+{\cal O}(1/r)$.  In this paper 
we
compute $\Psi_4^{l m}$ in the same way and at the same extraction
radius for all runs, so the ${\cal O}(1/r)$
effects should not significantly affect our comparisons of these waveforms.

\begin{figure}
\centerline{\includegraphics[width=6in]{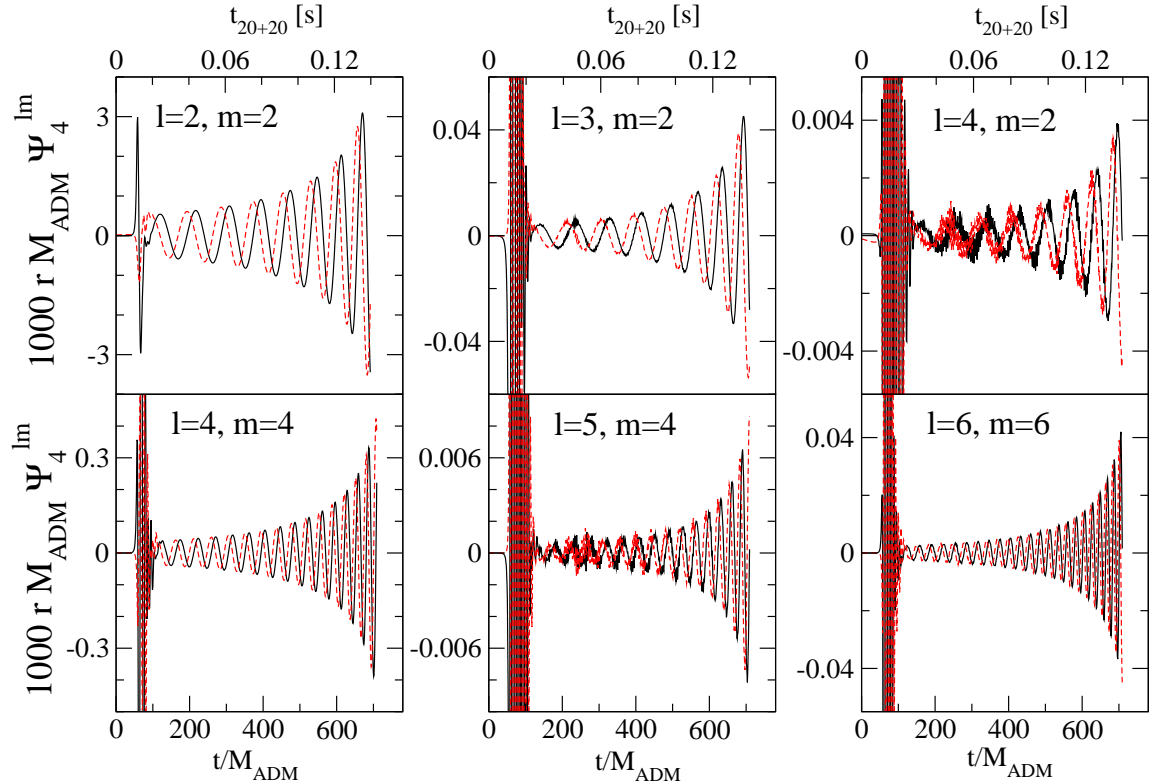}
}
\caption{\label{fig:Psi4}Waveforms for the F2 run.  Plotted are the six
  dominant $\Psi_4^{lm}$ coefficients, scaled by the factor
  $1000\,r\Eadm$. Solid lines represent the real parts and dashed
  lines the imaginary parts of $\Psi_4^{lm}$.  The time axes are labeled 
  in geometric
  units at the bottom, and in SI units for a 20+20 $M_\odot$ binary
  at the top.}
\end{figure}

Since our simulations use high spatial resolution all the way to
the outer boundary, the outgoing radiation is fully resolved
everywhere.  Therefore, we could extract waveforms at very
large radii.  The extracted wave signal lags the dynamics of
the binary by the light-travel time to the extraction radius, and
our evolutions currently fail shortly before merger.  So extracting
the wave signal at a very large radius would miss the most interesting part of
the waveform close to merger.  In order to retain most of the signal,
we compromise by extracting the radiation at an intermediate distance: 
$R/\Eadm=57$.
Figure~\ref{fig:Psi4} presents the dominant waveform coefficients 
$\Psi_4^{lm}$.  The $\Psi_4^{44}$ coefficient is about a factor of ten 
smaller than the largest coefficient, $\Psi_4^{22}$.  
The $\Psi_4^{32}$ and $\Psi_4^{66}$ coefficients are smaller by about
another order of magnitude; and the $\Psi_4^{42}$ and $\Psi_4^{54}$ 
coefficients have amplitudes that are only about $\sim 1/1000$ 
that of $\Psi_4^{22}$.

\subsection{Waveform comparisons}
\label{sec:4.4}

In this section we make a number of quantitative
comparisons between the waveforms produced by the evolution
of quasi-circular, QC, initial data and those produced by the
lower eccentricity, F, initial data.

\begin{figure}
\centerline{\includegraphics[width=4in]{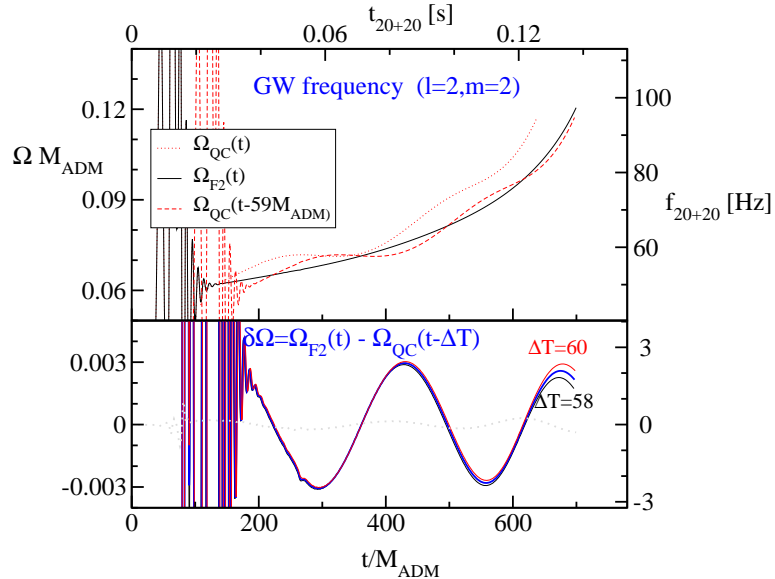}
}
\caption{\label{fig:OmegaGW}Frequency $\Omega_{22}$ of the gravitational 
  waves
  extracted from the phase of $\Psi_4^{22}$.  The left/bottom
  axes show geometric units, the right/top axes present SI-units for a
  20+20 $M_\odot$ binary.
  The dotted line in the lower panel represents the difference between the 
  E1 and F2 runs. }
\end{figure}

We can define a gravitational wave frequency associated with
$\Psi_4^{lm}$ by writing 
\begin{equation}
\Psi_4^{lm}=A_{lm}(t) e^{-i\phi_{lm}(t)}, 
\end{equation}
where $A_{lm}(t)$ is its (real) amplitude and $\phi_{lm}(t)$ its (real) phase.
The frequency, $\Omega_{lm}$, associated with $\Psi_4^{lm}$ is then 
defined as
\begin{equation}
\Omega_{lm}=\frac{d\phi_{lm}}{dt}.
\end{equation}
Figure~\ref{fig:OmegaGW} shows comparisons of the frequency of the dominant 
mode, $\Omega_{22}$, from the QC and the F runs.  This figure confirms
the basic picture that emerged from our discussion in Secs.~\ref{sec:4.1}
and \ref{sec:4.2}: a time offset $\Delta t$ must be used to compare the
QC and F runs properly; the QC run has an orbital
eccentricity which causes $\Omega_{22}$ to oscillate; and 
these oscillations are largely absent from the F run.
Indeed, apart from the factor of two difference between orbital
and the gravitational wave frequencies, the top panel of
Fig.~\ref{fig:OmegaGW} looks very much like Fig.~\ref{fig:OrbitalPhase}.
This indicates that our
coordinates are very well behaved---a feature that has also been observed
in other numerical simulations, e.g. Ref.~\cite{Pretorius2006}.

In order to make more detailed comparisons between the QC and the F 
waveforms, a phase offset $\Delta\phi$ in addition to the time offset 
$\Delta T$ must be taken into account. 
These offsets are used then to redefine
the waveform of the QC run:
\begin{equation}\label{eq:alphadeltatadjustment}
\tilde\Psi_{4\;\rm QC}^{lm}(t)\equiv e^{-im\Delta\phi}\; 
\Psi_{4\;\rm QC}^{lm}(t-\Delta T).
\end{equation}
The same time and phase offsets are used for all values of $l$
and $m$.  Note that $\Delta\phi$ and $\Delta T$ represent {\em differences}
between the QC and F evolutions.  These offsets
differ therefore from those often used in LIGO data analysis,
where offsets are used to set the time and orbital phase at which
a binary signal enters the LIGO band at 40Hz.

We now estimate the phase offset $\Delta\phi$ needed in
Eq.~(\ref{eq:alphadeltatadjustment}) to allow us to make direct comparisons
between the QC and the F2 waveforms.  We consider two effects: First,
the orbital phase of the time-shifted QC run differs from that of the
F2 run by the phase accumulated by the F2 run during the time $0\le
t\le \Delta T$.  Second, the orbital frequencies of the QC and F2 runs
differ, and this difference oscillates in time
(cf. the right panel of Fig.~\ref{fig:ProperSep}), 
so the orbital phase difference between the two runs also oscillates
in time.  
We take both of these effects into account: first, we
evaluate the time-dependent phase difference, $\Delta \phi(t)$, between the
waveforms of the time offset QC run, $\Psi_{4\,\mathrm{QC}}(t-\Delta t)$,
and the F run, $\Psi_{4\,\mathrm{F}}(t)$; second, we evaluate
the time average of this $\Delta\phi(t)$ to obtain 
$\Delta\phi\approx1.83$.
Using this value of $\Delta\phi$ leads to waveforms for the QC and F2
evolutions that agree as well as can be expected
in the presence of the other systematic errors, described below.

The two gravitational wave polarizations, $h_+(t)$ and $h_\times(t)$, are
the real functions related to $\Psi_4$ by
\begin{equation}
\Psi_4=\ddot{h}_+-i\ddot{h}_\times.
\end{equation}
Consequently, the $_{-2}Y_{lm}$ components of $h_+(t)$ and
$h_\times(t)$ can be obtained by the double time integral,
\begin{equation}\label{eq:time-integral}
h_{+}^{lm}(t)-i h_{\times}^{lm}(t)=\int_{t_i}^t 
d\tau \int_{t_i}^\tau d\tau' \Psi_{4}^{lm}(\tau')+C_{lm}+D_{lm}t.
\end{equation}
The constants $C_{lm}$ and
$D_{lm}$ account for the (unknown) values of $h$ and $\dot{h}$ at the
initial time $t_i$.  If the full waveform were known, they could be
determined either at very early times or at very late times
(i.e. after the merger and ringdown).
Since we do not have complete waveforms for the present evolutions,
we choose $C_{lm}$ and $D_{lm}$ that make the average and the first
moment of $h_{+\times}^{lm}(t)$ vanish:
\begin{equation}\label{eq:time-integral2}
\int_{t_1}^{t_2}d\tau\;h_{+\times}^{lm}(\tau)=0=\int_{t_1}^{t_2}d\tau
\;\tau\,h_{+\times}^{lm}(\tau).
\end{equation}
The integration interval $[t_1,t_2]=[160\Eadm,706\Eadm]$ is
chosen to be the largest interval (excluding the initial transient 
radiation burst) on which data is available for
both runs.

\begin{figure}
\centerline{\includegraphics[width=5.8in]{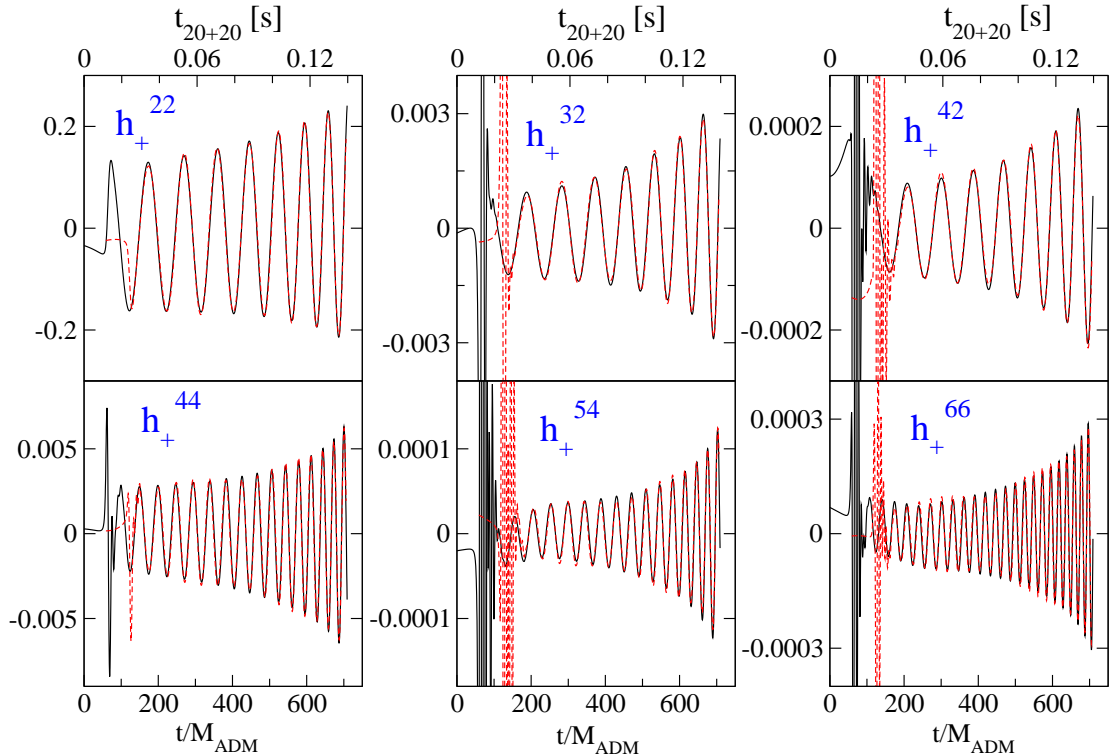}
}
\caption{\label{fig:h}Waveforms $h_{+}^{lm}$ (normalized by $r/\Eadm$) 
  for the
  six dominant $_{-2}Y^{lm}$ modes.  The solid lines represent
  evolution of the low-eccentricity initial data (run F2).  The dashed
  lines represent evolution of QC initial data time-shifted by
  $\Delta T=59\Eadm$ and phase-rotated by $\Delta\phi=1.83$.  The
  time axes are labeled in geometric units at the bottom and in
  SI-units for a 20+20 $M_\odot$ inspiral at the top.  }
\end{figure}

Figure~\ref{fig:h} shows the waveforms $h_+^{lm}$ for the evolution
F2 (solid lines) and QC (dashed lines).  To the eye, the
waveforms look essentially identical.  To quantify how well the two
waveforms match, we use simple overlap integrals in the time
domain:
\begin{equation}\label{eq:overlap}
\FittingFactor = \frac{\langle h_1,h_2\rangle}{||h_1||\; ||h_2||},
\end{equation}
where $\langle h_1,h_2\rangle\equiv\int_{t_1}^{t_2} dt\,h_1(t)h_2(t)$, and 
$||h||^2\equiv\langle h,h\rangle$.
The quantity $\FittingFactor$ gives the loss of signal to noise ratio
obtained by filtering waveform $h_1$ with waveform $h_2$.  We evaluate
the overlap integral in the time domain, rather than the frequency
domain, to allow us to truncate the waveforms easily to the
interval $[t_1,t_2]$ during which both waveforms are available.
During the evolutions presented here the gravitational-wave frequency
changes by only a factor of two, so our decision not to weight by the LIGO
noise spectrum should not change our results significantly for
frequencies near the minimum of the noise curve.
Furthermore, we evaluate $\mu$ directly for the different modes
$h_{+,\times}^{lm}$, rather than for specific observation directions.
This allows us to compare differences in
the higher order modes with smaller amplitudes, which would otherwise
be swamped by the dominant $l=m=2$ mode.

\begin{table}
\caption{\label{tab:overlap} Waveform overlaps between the
  low-eccentricity run F2 and quasi-circular run QC (computed from
  runs with medium and high resolution).  Each mode of QC
  has been time shifted and rotated by $\Delta T=59\Eadm$ and
  $\Delta\phi=1.83$.  These numbers are subject to additional systematic
  effects as discussed in the text.  }
\begin{indented}
\item[]
\begin{tabular}{c|cc|cc }
\br
& \multicolumn{2}{|c|}{high resolution} &
  \multicolumn{2}{|c}{medium resolution} \\
mode & 
$\FittingFactor(h_{+F}^{lm},\bar h_{QC+}^{lm})$  
& $\FittingFactor(h_{F \times}^{lm},\bar h_{QC\times }^{lm})$  &
$\FittingFactor(h_{+F}^{lm},\bar h_{QC+}^{lm})$  
& $\FittingFactor(h_{F \times}^{lm},\bar h_{QC\times }^{lm})$  \\
\mr
l=2, m=2 & $0.998$ & $0.998$  & 0.998 & 0.998  \\
l=3, m=2 & $0.997$ & $0.997$  & 0.997 & 0.998  \\
l=4, m=2 & $0.996$ & $0.997$  & 0.996 & 0.998  \\ 
l=4, m=4 & $0.991$ & $0.991$  & 0.993 & 0.993  \\
l=5, m=4 & $0.987$ & $0.979$  & 0.983 & 0.982  \\
l=6, m=6 & $0.981$ & $0.980$  & 0.986 & 0.982  \\
\br
\end{tabular}
\end{indented}
\end{table}

The overlaps between the QC and the F2 waveforms, obtained at
$\Delta T=59\Eadm$ and $\Delta\phi=1.83$, are summarized in
Table~\ref{tab:overlap}.  Both medium and high resolution overlaps
are given in Table~\ref{tab:overlap}, confirming that the overlaps
are not dominated by numerical errors.  We note, however, that the
medium resolution runs have more noise in the higher order modes at
early times; so we shortened the integration interval to
$[t_1,t_2]=[200\Eadm,706\Eadm]$ to avoid contamination in those
waveforms.  

The dominant uncertainty in the computed overlap $\mu$ arises because of
our uncertainties in the integration constants $C_{lm}$
and $D_{lm}$ in
Eqs.~(\ref{eq:time-integral}) and (\ref{eq:time-integral2}).  Because
the waveform has finite length, these constants are known
only to an accuracy of $\sim 1/N_{\rm cyc}$, where $N_{\rm cyc}$ is
the number of cycles in the waveform.  This error depends only
on the length of the evolution, and can only be reduced by longer evolutions,
not by higher resolution evolutions.  We show in~\ref{app:B}
(to lowest order in the uncertainties of $C_{lm}$
and $D_{lm}$) that the overlaps quoted in
Table~\ref{tab:overlap} are {\em upper} bounds.  We also
derive lower bounds for these overlaps there, which are smaller
than the values given in Table~\ref{tab:overlap} by about
$12/(\pi N_{\rm cyc})^2$.  So these lower bounds are
about 0.02 smaller than the Table~\ref{tab:overlap} values for the $m=2$
modes, and 0.002 smaller for 
the $m=6$ modes. 
This systematic uncertainty is much larger than the mismatch of the
waveforms for the $m=2$ modes, 
so maximizing the overlaps by varying $\Delta T$ and $\Delta\phi$
as independent parameters is not justified.

\section{Discussion}
\label{sec:5}

In this paper, we have extended the quasi-equilibrium initial-data
formalism to binary black holes with nonzero radial velocities.  We
have also used this formalism to construct initial data whose
evolution results in very low eccentricity orbits: about an order of
magnitude smaller than the orbits of quasi-circular initial data.

The main differences between evolutions of the quasi circular, QC, and 
the low eccentricity, F, initial
data sets are overall time and phase shifts: the QC initial data represents
the binary at a point closer to merger.  When we correct for these
shifts, the orbital trajectories of the black holes and the gravitational
waveforms they produce agree very well
between the two runs.  Various
parameters measured in the QC run (e.g.  orbital frequency or
proper separation) oscillate around the corresponding values from the F
run.  The gravitational wave phase oscillates as well, but no significant
coherent phase difference builds up during the five orbits
studied here. 
We find waveform overlaps between the 
high-eccentricity
and low-eccentricity runs of about 0.99.  Therefore it
appears that for the last five orbits before merger 
the differences between quasi-circular and low-eccentricity initial
data are not important for event detection in gravitational wave
detectors.  Longer evolutions (e.g. equal mass
binaries starting at larger separation, as well as unequal mass
binaries with a longer radiation reaction time scale) have more cycles
during which phase shifts could in principle accumulate.  However, orbital
eccentricity tends to decay during an inspiral and the orbital
eccentricity in quasi-circular data should decrease as the initial separation
increases, so longer evolutions are probably
less sensitive to the eccentricity in the initial data.
Thus we anticipate that the eccentricity of quasi-circular initial
data will not play a significant role when longer evolutions are
used for event detection,
but further study would be needed to confirm this.

Finally, we note that construction of low-eccentricity inspiral initial data
may be more difficult when the black holes carry generic spin.
The process described in this paper merely adjusts the orbital
parameters to obtain a trajectory without oscillations on the orbital
timescale.
For non-spinning equal-mass black holes sufficiently far
from merger, a non-oscillatory inspiral trajectory seems to be a
reasonable choice.  But if non-negligible spins are present, this is not
likely to be the case.  For  spins that are not aligned
with the orbital angular momentum, the approximate helical Killing
vector is lost, and there are likely to be a variety of oscillations 
on the orbital
time scale.  In these cases a more sophisticated model of the desired
circularized orbit will be needed before a procedure for adjusting
the orbital parameters to the appropriate values can be formulated.

\ack We thank Gregory Cook for inspiring discussions, including the
initial suggestion to add radial motion to quasi-circular initial
data.  We also thank Ilya Mandel for suggesting the comparison to
Peters' calculation~\cite{Peters1964}.  This work was supported in
part by grants from the Sherman Fairchild Foundation to Caltech and
Cornell, and from the Brinson Foundation to Caltech; by NSF grants
PHY-0099568, PHY-0244906, PHY-0601459, DMS-0553302 and NASA grants
NAG5-12834, NNG05GG52G at Caltech; and by NSF grants PHY-0312072,
PHY-0354631, and NASA grant NNG05GG51G at Cornell.  Some of the
simulations discussed here were produced with LIGO Laboratory
computing facilities. LIGO was constructed by the California Institute
of Technology and Massachusetts Institute of Technology with funding
from the National Science Foundation and operates under cooperative
agreement PHY-0107417. This paper has been assigned LIGO Document
Number LIGO-P060071-00-Z.

\appendix
\section{Quasi-equilibrium initial data in inertial coordinates}
\setcounter{section}{1}
\label{app:A}

In this appendix we show that $(\CF_{\rm co}, \Shift^i_{\rm
co}-\xirot, \Lapse_{\rm co})$, where $\xi^i_{\rm
rot}=(\mathbf{\OmegaOrbitID}\times \mathbf{r})^i$, is a solution to
the XCTS Eqs.~(\ref{eq:XCTSa})--(\ref{eq:XCTSc}) in asymptotically
inertial coordinates (with appropriately modified boundary conditions)
whenever $(\CF_{\rm co}, \Shift^i_{\rm co}, \Lapse_{\rm co})$ is a
solution in co-rotating coordinates.  We also show that this solution
leads to the same physical metric $\SMetric_{ij}$ and extrinsic
curvature $\ExCurv_{ij}$ as the original solution in co-rotating
coordinates.  The proof relies on three key observations: First, both
solutions are assumed to make the same choice of free data
Eqs.~(\ref{eq:utilde=0}) (\ref{eq:dtK=0}), (\ref{eq:ConfFlatness}),
and (\ref{eq:K=0}); second, the shift enters the XCTS equations and
the boundary conditions (almost) solely through the conformal Killing
operator, ${\CLong\Shift}^{ij}$; and third, $\xirot^i$ is a conformal
Killing vector, so $\CLong{\xirot}^{ij}=0$.  Hence the term
$-\xirot^i$ that is added to $\Shift^i_{\rm co}$ (mostly) drops out of
the equations.

We first show that the XCTS equations remain satisfied: Since
$\CLong{\xirot}^{ij}=0$, it follows from Eq.~(\ref{eq:A}) that $\tilde
A_{ij}$ is unchanged by the addition of $\xirot^i$.  So
Eqs.~(\ref{eq:XCTSa}) and~(\ref{eq:XCTSb}) remain satisfied.  The only
other shift containing term in Eq.~(\ref{eq:XCTSc}) is
$\Shift^i\partial_i\TrExCurv$, which vanishes because
$\partial_i\TrExCurv=0$ from the choice of free data ($K=0$) in
Eq.~(\ref{eq:K=0}); so Eq.~(\ref{eq:XCTSc}) also remains satisfied.

We turn next to the boundary conditions: The boundary conditions used
for the co-rotating coordinate representation of the XCTS equations
are Eqs.~(\ref{eq:AsympFlat})--(\ref{eq:ZeroShear}) and
(\ref{eq:LapseBC}), while those used for the inertial frame
representation are the same, except Eqs.~(\ref{eq:OuterBC}) and
(\ref{eq:dtAH=0}) are replaced by Eqs.~(\ref{eq:QC-inertial1}) and
(\ref{eq:QC-inertial2}).  The boundary conditions,
Eqs.~(\ref{eq:AsympFlat}) and (\ref{eq:LapseBC}), depend only on $\CF$
and $\Lapse$ and therefore remain satisfied.  The apparent horizon
boundary condition, Eq.~(\ref{eq:AH}), implies the boundary
condition on the conformal factor Eq.~(\ref{eq:AH-BC}), which
is unchanged since $\CLong{\xirot}^{ij}=0$; and the new outer
boundary condition, Eq.~(\ref{eq:QC-inertial1}), also holds because 
$\Shift^i_{\rm co}$ satisfies Eq.~(\ref{eq:OuterBC}).

The only remaining boundary conditions then are 
Eqs.~(\ref{eq:QC-inertial2}) and
(\ref{eq:ZeroShear}).  Because $\NullExpansion=0$ and
$\NullShear_{ij}=0$, the null surface generated by $\NullNormal^\mu$
coincides with the world tube of the apparent horizons, ${\cal S}_{\rm
  AH}$.  The normal to this null surface is $\NullNormal^\mu$, because
$\NullNormal^\mu$ is normal to ${\cal S}$ by construction, and because
$\NullNormal^\mu\NullNormal_\mu=0$.  Therefore, in order for
$\partial_t+\xirot^i\partial_i$ to be tangent to ${\cal S}_{\rm AH}$,
as required by the boundary condition Eq.~(\ref{eq:QC-inertial2}),
it must be orthogonal to $\NullNormal^\mu$.  The vector
$\partial_t+\xirot^i\partial_i$ has components $\Lapse
n^\mu+\Shift^\mu+\xirot^\mu$, where $\Shift^\mu=[0,\Shift^i]$ and
$\xirot^\mu=[0,\xirot^i]$.  Using $k^\mu=(n^\mu+s^\mu)/\sqrt{2}$,
it follows that
\begin{equation}
0=(\partial_t+\xirot^i\partial_i)\cdot\NullNormal
=\frac{1}{\sqrt{2}}\left[-\Lapse+(\Shift^i+\xirot^i)\SSpatialNormal_i\right].
\end{equation}
This condition implies 
\begin{equation}
\label{eq:ShiftBC-inertial2}
\Shift^i=\Lapse\SSpatialNormal^i-\xirot^i+\zeta^i\quad\mbox{on ${\cal S}$},
\end{equation}  
with $\zeta^i\SSpatialNormal_i=0$,
i.e., Eq.~(\ref{eq:ShiftBC-inertial}) in the main
text.  So the boundary condition Eq.~(\ref{eq:QC-inertial2}) 
is satisfied because $\beta^i_{\rm co}=\Lapse\SSpatialNormal^i+\zeta^i$ 
satisfies Eq.~(\ref{eq:ShiftBC}).

The vector $\zeta^i$ that appears in Eq.~(\ref{eq:ShiftBC-inertial2})
is further constrained by the shear boundary condition,
Eq.~(\ref{eq:ZeroShear}), which we consider next.  
The shear $\NullShear_{ij}$ is defined as
\begin{equation}
\NullShear_{\mu\nu}=\perp_{\mu\nu}^{\;\;\;\;\;\rho\sigma}
\FourCD_\rho k_\sigma,
\end{equation}
where $\perp_{\mu\nu}^{\;\;\;\;\;\rho\sigma}
=\STwoMetric_\mu{}^{(\rho}\STwoMetric_\nu{}^{\sigma)}
-\frac{1}{2}\STwoMetric_{\mu\nu}\STwoMetric^{\rho\sigma}$.
Substituting Eq.~(\ref{eq:NullNormal}) into this expression, and
subsequently using Eqs.~(\ref{eq:DefExCurv}),~(\ref{eq:K-split}),
and~(\ref{eq:A}) results in
\begin{equation}
\sqrt{2}\NullShear_{ij}=
-\frac{1}{2\Lapse}\perp_{ij}^{\;\;\;kl} \left[\SLong{\Shift}_{kl}
-\CF^4\dtCMetric_{kl}\right]+\perp_{ij}^{\;\;\;kl}\SCD_k\SSpatialNormal_l.
\end{equation}
For any
vector field $v^i$ decomposed into normal and tangential parts,
$v^i=v^m\SSpatialNormal_m\, \SSpatialNormal^i+v_{||}^i$, it follows that
\begin{equation}
\perp_{ij}^{\;\;\;kl}\SLong{v}_{kl}=\SLong{_{\cal S}v_{||}}_{ij}+2v^m\SSpatialNormal_m\, 
\perp_{ij}^{\;\;\;kl}\SCD_k\SSpatialNormal_l.
\end{equation}
Using this identity and Eq.~(\ref{eq:ShiftBC-inertial}), the shear can be
rewritten as
\begin{equation}
\sqrt{2}\NullShear_{ij}=\frac{1}{2\Lapse}\perp_{ij}^{\;\;\;kl}\left[\SLong{\xirot}_{kl}+\CF^4\dtCMetric_{kl}\right]-\frac{1}{2\Lapse}\SLong{_{\cal S}\zeta}_{ij}.
\end{equation}
Once more, $\xirot^i$ drops out because it is a conformal Killing vector.
Also, since $\dtCMetric_{ij}=0$ by Eq.~(\ref{eq:utilde=0}), we find
that the shear vanishes iff $\zeta^i$ is a conformal Killing vector within the
2-surface ${\cal S}$:
\begin{equation}
\NullShear_{ij}=0\quad\Leftrightarrow\quad
 0=\SLong{_{\cal S}\zeta}^{ij}.
\end{equation}
Equation~(\ref{eq:Lzeta=0}) now follows from the identity $\SLong{_{\cal
    S}\zeta}^{ij}=\CF^{-4}\CLong{_{\cal S}\zeta}^{ij}$.  This implies
then that the boundary condition Eq.~(\ref{eq:ZeroShear}) is satisfied
since it is assumed to be satisfied in the co-rotating case.

Finally, we note that the physical metric $\SMetric_{ij}$ and extrinsic
curvature $\ExCurv_{ij}$ produced by the inertial frame version
of the problem are identical to those of the original co-rotating
frame version.  The conformal metric and conformal factor are identical
in the two versions, so the physical metrics are identical trivially 
from Eq.~(\ref{eq:CMetric}).  Since $\xirot^i$ is a conformal Killing
vector, it follows that $A_{ij}$ is identical
from Eq.~(\ref{eq:A}); so it follows from Eq.~(\ref{eq:K-split})
(with $K=0$) that the extrinsic curvatures are identical as well.

\section{Errors caused by finite-length waveforms}
\label{app:B}

The error in the waveform overlaps caused by the uncertainty in the
integration constants can be determined as follows: Denote
our numerically computed
waveforms by $h_x+\varepsilon_x$, where $h_x$ stands for
the unknown ``true'' waveform obtained with the correct values of the
integration constants, and $\varepsilon_x$ represents the error
introduced by computing these constants with a truncated
waveform.  The label $x$ stands for either $F$ or $QC$.

The quantity of interest is the overlap between the ``true'' waveforms,
\begin{equation}
\FittingFactor(h_F,h_{QC})=\frac{\langle h_F,h_{QC}\rangle}
{||h_F||\;||h_{QC}||},
\end{equation}
where $\langle h_1,h_2 \rangle\equiv\int_{t_1}^{t_2}h_1(t)h_2(t)dt$, and
$||h||^2\equiv \langle h,h\rangle$.  
The errors $\varepsilon_x$ are those caused
by the uncertainty in the constants $C_{lm}$ and $D_{lm}$ in
Eq.~(\ref{eq:time-integral}), and the $\varepsilon_x$
are therefore {\em linear} functions of time.
Furthermore, choosing the integration constants by
Eq.~(\ref{eq:time-integral2}) makes the numerical waveforms
$h_x+\varepsilon_{x}$ {\em orthogonal} to functions linear in time,
so that $\langle h_x+\varepsilon_x,\varepsilon_y\rangle=0$, 
where $x,y\in\{F, QC\}$.
Using this result, and neglecting terms of order 
${\cal O}(\varepsilon^3)$, one finds
\begin{eqnarray}\label{eq:Ba}
\fl
\FittingFactor(h_F+\varepsilon_F,h_{QC}+\varepsilon_{QC})
&=&\FittingFactor(h_F,h_{QC})\nonumber\\
&&\,\,\,+\FittingFactor(h_F,h_{QC})
\left(\frac{||\varepsilon_F||^2}{2||h_F||^2}
+\frac{||\varepsilon_{QC}||^2}{2||h_{QC}||^2}
-\frac{\langle\varepsilon_F,\varepsilon_{QC}\rangle}
{||h_F||\;||h_{QC}||}\right).
\end{eqnarray}
It is straightforward to
show that $\FittingFactor(h_f,h_{QC})=1-{\cal O}(\delta h^2)$
where $\delta h=h_F-h_{QC}$. Therefore, replacing
$\FittingFactor(h_F,h_{QC})\to 1$ in the last term of
Eq.~(\ref{eq:Ba}) changes the result only by terms of order ${\cal O}(\delta
h^2\,\varepsilon_x^2)$.  Furthermore, replacing
$||h_{QC}||\to||h_F||$ in the denominators of Eq.~(\ref{eq:Ba}) 
affects the result only by terms of order ${\cal O}(\delta h\,\varepsilon^2)$.
Neglecting both of these higher order contributions, we find

\begin{equation}\label{eq:Bb}
\FittingFactor(h_F+\varepsilon_F,h_{QC}+\varepsilon_{QC})
=\FittingFactor(h_F,h_{QC})
+\frac{||\varepsilon_F-\varepsilon_{QC}||^2}{2||h_F||^2}.
\end{equation}
Because the last term is non-negative, the ``true'' overlap
$\FittingFactor(h_F,h_{QC})$ is always smaller than the numerically computed
overlap $\FittingFactor(h_F+\varepsilon_F,h_{QC}+\varepsilon_{QC})$.
Using the triangle inequality, we can
bound the last term in Eq.~(\ref{eq:Bb}) by the error
$||\varepsilon_x||^2/||h_x||^2$ in either the F or the QC waveform:
\begin{equation}
\frac{||\varepsilon_F-\varepsilon_Q||^2}{2||h_F||^2}\le
\frac{(||\varepsilon_F||+||\varepsilon_{QC}||)^2}{2||h_F||^2}
\approx 2 \frac{||\varepsilon_x||^2}{||h_x||}.
\end{equation}

Finally, we estimate $||\varepsilon_x||^2/||h_x||^2$ by applying
Eqs.~(\ref{eq:time-integral}) and (\ref{eq:time-integral2}) to a pure
sine-wave: $h(t)=\sin(t)$.  It is straightforward to
evaluate the integrals in Eq.~(\ref{eq:time-integral2})
for this simple case, giving
the bound $||\varepsilon||^2/||h||^2\le 6/(\pi N_{\rm cyc})^2$, where
$N_{\rm cyc}=(t_2-t_1)/(2\pi)$ is the number of cycles in the interval
$[t_1,t_2]$.  Therefore, we arrive at the bounds
\begin{equation}\fl
\FittingFactor(h_F+\varepsilon_F,h_{QC}+\varepsilon_{QC})\ge 
\FittingFactor(h_F,h_{QC})\gtrsim 
\FittingFactor(h_F+\varepsilon_F,h_{QC}+\varepsilon_{QC})-\frac{12}{\pi^2N_{\rm cyc}^2},
\end{equation}
as mentioned in the main text.\\ \\ \\


\end{document}